\documentclass[a4paper,fleqn,usenatbib]{mnras}

\usepackage{gensymb}
\usepackage{graphicx}
\graphicspath{{figs/}}
\usepackage{verbatim}
\usepackage{amssymb}
\usepackage{lineno}
\usepackage{wasysym}
\usepackage{amssymb}
\usepackage{xcolor}
\usepackage{soul}
\usepackage{amstext}
\usepackage{array}

\newcommand{\kms}{km~s$^{-1}$}

\newcommand{\mgii}{\ion{Mg}{ii}}
\newcommand{\caii}{\ion{Ca}{ii}}
\newcommand{\mgi}{\ion{Mg}{i}}
\newcommand{\hi}{\ion{H}{i}}
\newcommand{\feii}{\ion{Fe}{ii}}
\newcommand{\oii}{[\ion{O}{ii}]}

\newcommand{\ciii}{\ion{C}{iii}]}
\newcommand{\nhi}{$N($\ion{H}{i}$)$}

\newcommand{\rcs}{RCS2032727$-$132623}
\newcommand{\pks}{PSZ1 G311.65--18.48}

\newcommand{\zabs}{0.73379}

\newcommand{\msun}{M$_{\astrosun}$}
\newcommand{\msunm}{{\rm M}_{\astrosun}}

\newcommand{\rfemg}{$\cal{R}^{\rm FeII}_{\rm MgII}$}
\newcommand{\rmgimg}{$\cal{R}^{\rm MgI}_{\rm MgII}$}

\newcommand{\PAne}{72$\degree$}
\newcommand{\PAsky}{52.3$\degree$}
\newcommand{\PAsw}{42$\degree$}
\newcommand{\netot}{2.0}  
\newcommand{\swtot}{3.0}  
\newcommand{\skytot}{1.17}  

\newcommand{\galpak}{{\sc Galpak}}
\newcommand{\lenstool}{{\sc Lenstool}}
\newcommand{\drizzlepac}{{\sc Drizzlepac}}

\title[Slicing the CGM at $z=0.7$]{Slicing the cool circumgalactic medium along the major-axis of
a star-forming galaxy at $z=0.7$}

\author[Lopez et al.]{S.~Lopez,$^{1}$ 
N.~Tejos,$^{2}$ 
L.~F.~Barrientos,$^{3}$
C.~Ledoux,$^{4}$
K.~Sharon,$^{5}$
A.~Katsianis,$^{1,6,7}$
\newauthor
M.~K.~Florian,$^{8}$
E.~Rivera-Thorsen,$^{9}$
M.~B.~Bayliss,$^{10,11}$
H.~Dahle,$^{12}$
\newauthor
A.~Fernandez-Figueroa,$^{1}$
M.~D.~Gladders,$^{13}$
M.~Gronke,$^{14}$
M.~Hamel,$^{1}$
I.~Pessa$^{15}$
\newauthor
and
J.~R.~Rigby$^{16}$
\\
$^{1}$ Departamento de Astronom\'ia, Universidad de Chile, Casilla 36-D,
Santiago, Chile. E-mail: slopez@das.uchile.cl\\
$^{2}$ Instituto de F\'\i sica, Pontificia Universidad Cat\'olica de
Valpara\'\i so, Casilla 4059, Valpara\'\i so, Chile. E-mail: nicolas.tejos@pucv.cl\\
$^{3}$ Instituto de Astrof\'\i sica, Pontificia Universidad Cat\'olica de Chile, Casilla 306, Santiago, Chile\\
$^{4}$ European Southern Observatory, Alonso de C\'ordova 3107, Vitacura, Casilla 19001, Santiago, Chile\\
$^{5}$ Department of Astronomy, University of Michigan,  Ann Arbor, MI 48109, USA\\
$^{6}$ Tsung-Dao Lee Institute, Shanghai Jiao Tong University, Shanghai 200240, China\\
$^{7}$ Department of Astronomy, Shanghai Key Laboratory for Particle Physics
    and Cosmology, Shanghai Jiao Tong University, \\
\ \ \ Shanghai 200240, China\\
$^{8}$ Observational Cosmology Lab, Goddard Space Flight Center, Code 665, Greenbelt, MD 20771, USA\\
$^{9}$ Institute of Theoretical Astrophysics, University of Oslo, Postboks 1029, 0315 Oslo, Norway\\
$^{10}$ Kavli Institute for Astrophysics \& Space Research, Massachusetts
Institute of Technology, 77 Massachusetts Avenue, \\
\ \ \ Cambridge, MA 02139, USA\\
$^{11}$ Department of Physics, University of Cincinnati, Cincinnati, OH 45221, USA\\
$^{12}$ Institute of Theoretical Astrophysics, University of Oslo, P.O. Box 1029, Blindern, NO-0315 Oslo, Norway\\
$^{13}$ Department of Astronomy \& Astrophysics and Kavli Institute for
Cosmological Physics, University of Chicago, 5640 South Ellis Avenue, \\
\ \ \  Chicago, IL 60637, USA\\
$^{14}$ Department of Physics, University of California, Santa Barbara, CA 93106, USA\\
$^{15}$ Max-Planck-Institut f\"ur Astronomie, Königstuhl 17, D-69117 Heidelberg, Germany\\
$^{16}$ Observational Cosmology Lab, NASA Goddard Space Flight Center, Greenbelt, MD 20771, USA
}

\begin{document}
\label{firstpage}
\pagerange{\pageref{firstpage}--\pageref{lastpage}}
\maketitle

\begin{abstract}
We present spatially-resolved echelle spectroscopy of an intervening
\mgii-\feii-\mgi\ absorption-line system detected at $z_{\rm abs}=\zabs$
toward the giant gravitational arc \pks.  The absorbing gas is
associated to an inclined disk-like star-forming galaxy, whose major axis is
aligned with the two arc-segments reported here.  We probe in absorption the
galaxy's extended disk continuously, at $\approx 3$\,kpc sampling, from its
inner region out to $15\times$ the optical radius. We detect strong
($W_0^{2796}>0.3$\,\AA) coherent absorption along $13$ independent positions
at impact parameters $D=0$--$29$\,kpc on one side of the galaxy, and no
absorption at $D=28$--$57$\,kpc on the opposite side (all de-lensed distances at
$z_{\rm abs}$).  We show that: (1) the gas distribution is anisotropic; (2)
$W_0^{2796}$, $W_0^{2600}$, $W_0^{2852}$, and the ratio $W_0^{2600}\!/W_0^{2796}$, all
anti-correlate with $D$; (3) the $W_0^{2796}$-$D$ relation is not cuspy and
exhibits significantly less scatter than the quasar-absorber statistics; (4)
the absorbing gas is co-rotating with the galaxy out to $D \lesssim 20$\,kpc,
resembling a `flat' rotation curve, but at $D\gtrsim 20$\,kpc velocities {\it
  decline} below the expectations from a 3D disk-model extrapolated from the
nebular \oii\ emission.  These signatures constitute unambiguous evidence for
rotating extra-planar diffuse gas, possibly also undergoing enriched accretion
at its edge.  Arguably, we are witnessing some of the long-sought processes of
the baryon cycle in a single distant galaxy expected to be representative of
such phenomena.
\end{abstract}


\section{Introduction} 
\label{introduction}

Models and simulations that describe the various components and scales of the
baryon cycle around galaxies remain to be tested observationally. Such a task
poses a serious challenge, though, as most of the `action' occurs in the
diffuse circum-galactic medium (CGM), i.e., at several optical radii from the
host galaxy scales ~\citep[e.g.,][]{Tumlinson2017}.  Traditionally,
observations of the CGM at $10$--$100$\,kpc scales have been based on the
absorption it imprints on background sources, primarily quasars
\citep[e.g.,][and references
  therein]{Nielsen2013cat,Prochaska2017,Tumlinson2017,Chen2017} but also
galaxies \citep{Steidel2010,Diamond-Stanic2016,Rubin2018}, including the
absorbing galaxy itself \citep{Martin2005,Martin2012,Kornei2012}. Such
techniques have yielded a plethora of observational constraints and evidence
for a connection between a galaxy's properties and its CGM.

Galaxies studied through these methods, nevertheless, are probed by single
pencil beams; therefore, to draw any conclusions that involve the spatial
dependence of an observable requires averaging absorber properties
\citep{Chen2010,Nielsen2013} or stacking spectra of the background sources
\citep{Steidel2010,Bordoloi2011,Rubin2018a,Rubin2018b}. A complementary
workaround is to use multiple sight-lines through individual
galaxies. Depending on the scales, the background sources can be binary or
chance quasar groups \citep{Martin2010,Bowen2016} or else lensed quasars
\citep{Smette1992,Lopez1999,Lopez2005,Lopez2007,Rauch2001,Ellison2004,Chen2014,Zahedy2016}.
Despite the paucity of the latter, lensed sources are able to resolve the CGM
of intervening galaxies on kpc scales, albeit at a sparse sampling. More
recently, \citet{Lopez2018} have shown that the spatial sampling can be
greatly enhanced by using giant gravitational arcs. Comparatively, these giant
arcs are very extended \citep[e.g.,][]{Sharon2019} and thus can probe the
gaseous halo of {\it individual} galaxies on scales of $1$--$100$\,kpc at a
{\it continuous} sampling, nicely matching typical CGM scales. Such an
experimental setup, therefore, removes potential biases introduced by
averaging a variety of absorbing galaxies.

Following on our first tomographic study of the cool CGM around a star-forming
group of galaxies at $z\approx 1$ \citep[][hereafter `Paper I']{Lopez2018}, we
here present spatially-resolved spectroscopy of a second giant gravitational
arc. We pool together echelle and integral-field (IFU) spectroscopy of the
brightest known gravitational arc to date, found around the cluster
\pks~\citep[a.k.a. the `Sunburst
  Arc';][]{Dahle2016,Rivera-Thorsen2017,Rivera-Thorsen2019,Chisholm2019}.
We apply our technique to study the spatial extent and kinematics of an
intervening \mgii-\feii-\mgi\ absorption-line system at $z=\zabs$.  Due to a
serendipitous arc/absorber geometrical projection on the sky, we are able to
spatially resolve the system all along the major axis of a host galaxy that
may be exemplary of the absorber population at these intermediate redshifts.

The paper is structured as follows. In Section~\ref{sec:data}, we present the
observations and describe the different datasets. In
Section~\ref{sec_lens_model}, we describe the reconstructed absorber plane and
assess the meaning of the absorption signal. In Section~\ref{sec:G1}, we
present the emission properties of the identified absorbing galaxy.  In
Section~\ref{sec:abs}, we provide the main analysis and results on the line
strength and kinematics of the absorbing gas. We discuss our results in
Section~\ref{sec:discussion} and present our summary and conclusions in
Section~\ref{sec:summary}. Details on data reduction and models are provided
in an Appendix. Throughout the paper, we use a $\Lambda$CDM cosmology with the
following cosmological parameters:
$H_0=70$\,km\,s$^{-1}$\,Mpc$^{-1}$, 
$\Omega_m=0.3$, and 
$\Omega_{\Lambda} =0.7$.

\section{Observations and data reduction}\label{sec:data}

\subsection{Experimental setup}

\begin{figure}
\includegraphics[width=\columnwidth,angle=0]{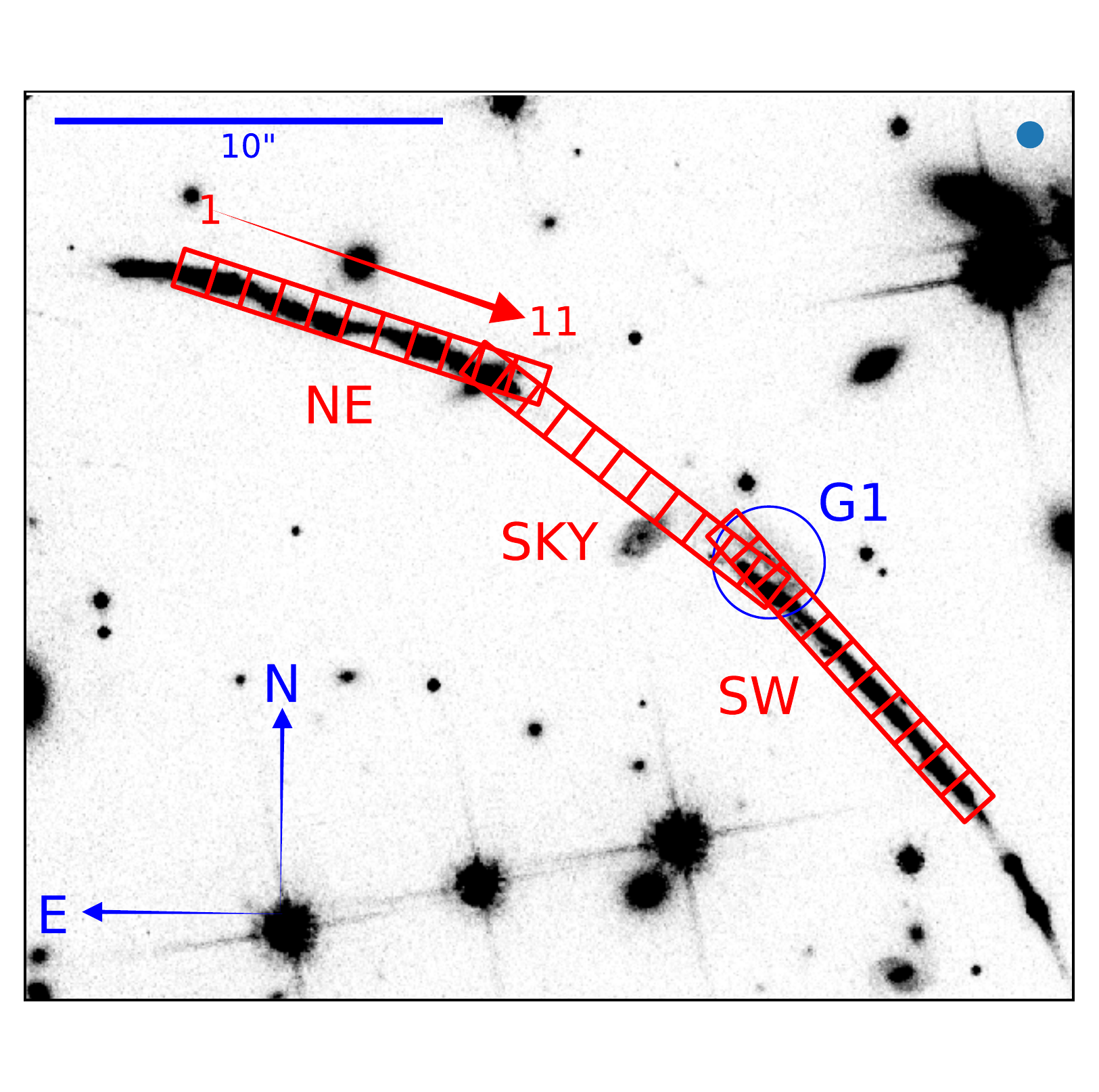} 
\caption{{\it HST}/ACS F814W-band image of the northern arc segments around \pks. The
$3$ MagE slits (`NE', `SKY', and `SW') are indicated in red, along with our definition of
`pseudo-spaxels', and their numbering (for clarity only shown for the NE slit;
  see~\S~\ref{sec:mage_data}). The slit widths are of $1$\arcsec, and their
  lengths are of $10$\arcsec; we have divided each of them into  $11$
  pseudo-spaxels of $1.0$\arcsec$\times0.9$\arcsec each. 
The position of the absorbing galaxy (G1) is encircled in blue. The ground-based
observations were performed under a seeing of $0.7"$ (represented by the beam-size
symbol in the top-right corner). 
\label{fig_FOV}}
\end{figure}

\begin{table*}
\caption{Summary of Magellan/MagE observations}\label{tab:obs}
\centering
\begin{tabular}{ccccccr}
\hline
\hline
Slit & PA & \multicolumn{2}{c}{Exposure Time}   & Airmass & Seeing &\multicolumn{1}{c}{Blind Offsets} \\
     & (degrees) & Individual (s)& Total (h)&    &  (\arcsec) &  \multicolumn{1}{c}{(\arcsec)} \\ 
(1) & (2) & (3) & (4) & (5) & (6) & \multicolumn{1}{c}{(7)} \\
     \hline
 SW  &  42.0   & $2700+3600+4500$ & $3.0$& $1.6$--$1.7$ & $0.6$--$0.7$&4.19(E), 10.22(S) \\
 SKY &  52.3 & $4200$ & $1.2$ &  $1.7$ & $0.7$--$0.9$ & 9.64(E), 5.39(S)\\
 NE  &  72.0  & $3600+3600$ & $2.0$& $1.5$--$1.6$  & $0.6$-$0.7$ & 16.51(E), 1.47(S)\\
\hline
\end{tabular}\\
\vspace{1ex}
\begin{minipage}{0.72\textwidth}
\raggedright {\bf Notes:} 
(1) Slit name (see Fig.~\ref{fig_FOV});
(2) Position angle of slit;
(3) Individual exposure times;
(4) Total exposure times;
(5) Airmass of the observations;
(6) Typical seeing FWHM of the observations;
(7) Acquisition blind offsets to the East (E) and South (S) from reference star at celestial coordinate (J2000) R.A.$=$ 15h\,50m\,00s and
Dec. $=-78$\degree\,10m\,57s.
\vspace{2ex}
\end{minipage}
\end{table*}

\begin{figure}
\includegraphics[width=\columnwidth,angle=0]{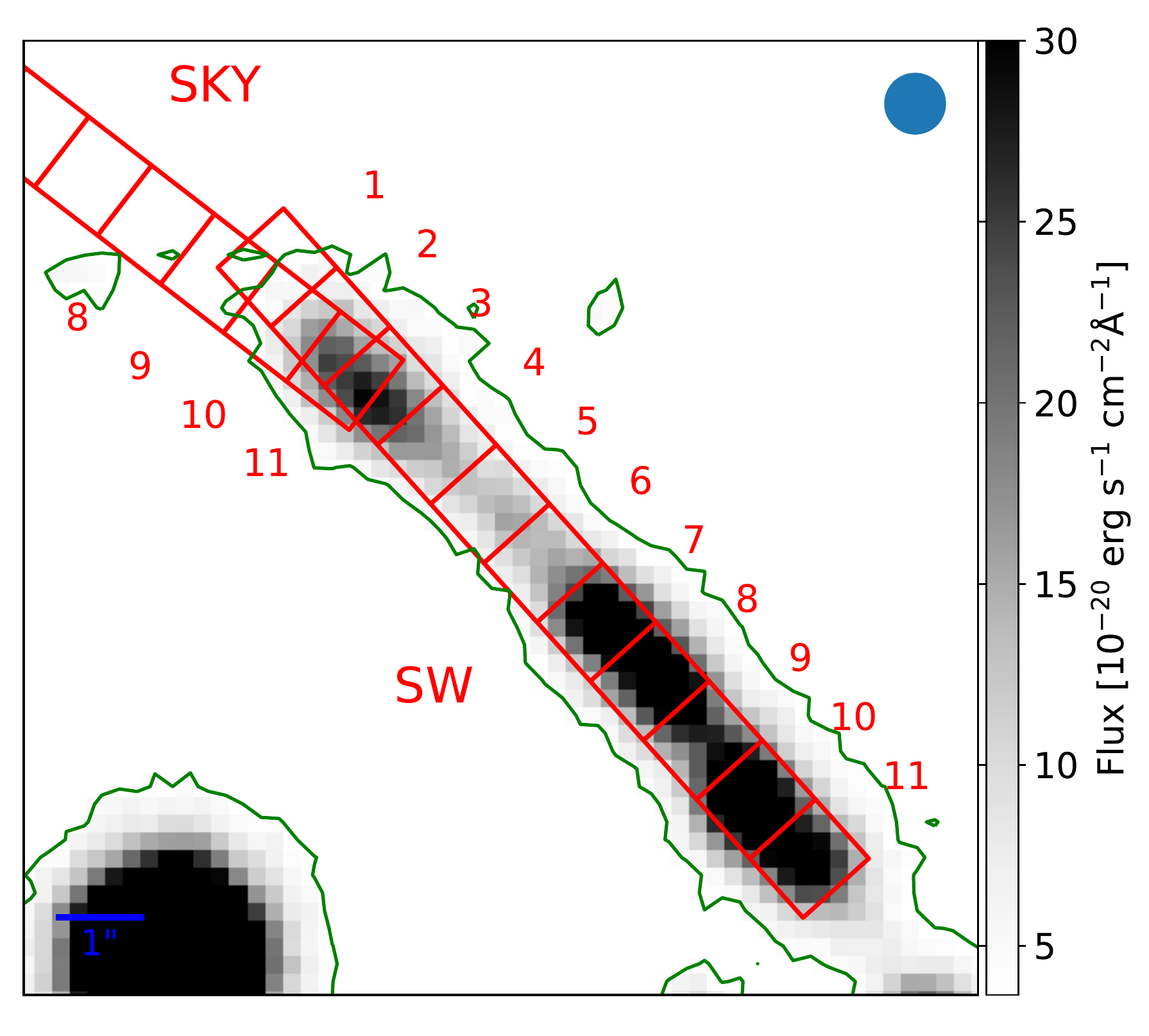} 
\caption{Zoom-in into the SW segment showing a MUSE image centered at the
  continuum around \mgii\ absorption at $\sim 4848$ \AA. The MagE `SW' and
  `SKY' slits with their corresponding pseudo-spaxels (\S~\ref{sec:mage_data})
  are shown in red.  The blue circle indicates the seeing FWHM. The green
  contours indicate a flux level of $5\sigma$ above the sky level. Since the
  observing conditions during the MUSE and the MagE observations were quite
  similar (e.g., dark nights, seeing $\approx 0\farcs7$), such contours show
  that SW pseudo-spaxels \#2 to \#11 and SKY pseudo-spaxels \#10 to \#11 were
  fully illuminated by the source, while SW \#1 was only partially illuminated
  by the source.  Using the same method, all NE pseudo-spaxels appear to be
  illuminated by the source (not shown here).
\label{fig_slit_sky}}  
\end{figure}

\pks\ extends over $\approx 60\arcsec$ on the sky (Fig.~\ref{fig_FOV}) and
results from the lensing of a $z=2.369$ star-forming galaxy by a cluster at
$z=0.443$ \citep{Dahle2016}.  According to archival VLT/MUSE data, an
intervening \mgii\ absorption-line system at $z = \zabs$ appears in the
spectra of one of northernmost segments of the arc. The same data reveal
nebular \oii\ emission at the same redshift from a nearby galaxy, which we
consider to be the absorbing galaxy (hereafter referred to as `G1').  To
thoroughly study this system, in this paper we exploit three independent
datasets: (1) medium-resolution IFU data obtained with VLT/MUSE, which we use
to constrain the emission-line properties of G1; (2) {\it Hubble Space
  Telescope} ({\it HST}) imaging, which we largely use to (a) build the lens
model needed to reconstruct the absorber plane, and (b) constrain the overall
properties of G1 based on its continuum emission; and (3) medium-resolution
echelle spectra obtained with Magellan/MagE, which we use to constrain the
absorption-line properties of the gas.

\subsection{VLT/MUSE}\label{sec:muse_data}

We retrieved MUSE observations of \pks\ from the ESO archive (ESO program 297.A-5012(A); PI Aghanim).
The field comprising the arc segments shown in Fig.~\ref{fig_FOV} 
was observed in wide-field mode 
for a total of $2966$\,s on the night of May 13th, 2016 under good seeing conditions ($0\farcs7$). 
We reduced the raw data using the MUSE pipeline v1.6.4 available in {\sc Esoreflex}. The sky subtraction was improved using the Zurich Atmospheric Purge (ZAP v1.0) algorithm.  We applied a small offset to the {\it HST} and MUSE fields to take them to a common astrometric system using as a reference a single star near G1.  
The spectra cover the wavelength range 
$4\,750$--$9\,300$ \AA\ at a resolving power $R\approx 2\,100$.  The exposure time resulted in a  S/N that is adequate to constrain
the emission-line properties of G1, but not enough for the absorption-line analysis, given the MUSE spectral resolution.

\subsection{HST/ACS}\label{sec:hst_data}

{\it HST} observations of \pks\ were conducted on February 21st to 22nd, 2018,
and September 2nd, 2018 using the F814W filter of ACS (GO15101; PI Dahle) and
the F160W filter of the IR channel of WFC3 (GO15337; PI Bayliss)
respectively. F814W observations consist of 8 dithered exposures acquired over
two orbits, totaling $5280$\,s.  F160W observations were conducted in one
orbit, using three dithered pointings totaling $1359$\,s.

These data were reduced using the \drizzlepac\ software
package.\footnote{\url{drizzlepac.stsci.edu}} Images were drizzled to a
$0.03$\arcsec  per pixel grid using the routine {\tt astrodrizzle} with a ``drop
size" (final$\_$pixfrac) of 0.8 using a Gaussian kernel.  Where necessary,
images were aligned using the routine tweakreg, before ultimately being
drizzled onto a common reference grid with north up.

\subsection{Magellan/MagE}\label{sec:mage_data}

Spectroscopically, Magellan/MagE greatly outperforms MUSE in terms of blue
coverage and resolving power; hence, {\it these observations are central to
  the present study}. Here we provide a concise description of the
observations (see Table~\ref{tab:obs} for a summary).  More details on the
observations and data reduction are presented in the
Appendix~\ref{sec:mage_appendix}.

We observed the two northernmost segments in \pks\ during dark-time on the
first half-nights of July 20th and 21st, 2017 (program CN2017B-57, PI
Tejos). The weather conditions varied but the seeing was good
($0\farcs6-0\farcs7$) and steady.

With the idea of mimicking integral-field observations, we placed three
$1$\arcsec $\times 10$\arcsec\ slits (referred to as `NE', `SKY' and `SW')
along the two arc segments (see Fig.~\ref{fig_FOV}) using blind offsets. The
`SKY' slit was placed in a way that the northernmost/southernmost extreme of
the slit has light contribution from the North-East/South-West arc segments,
respectively, while the inner part is dominated by the actual background sky
signal.  Thus, the `SKY' slit provides not only a reference sky spectrum for
the `NE' and `SW' slits (both completely covered by the extended emission of
the arc at seeing $0\farcs7$; see Fig.~\ref{fig_slit_sky}), but it also
provides independent arc signal at the closest impact parameters to G1 in each
arc segment.

\begin{figure}
\centering
\includegraphics[width=\columnwidth,angle=0]{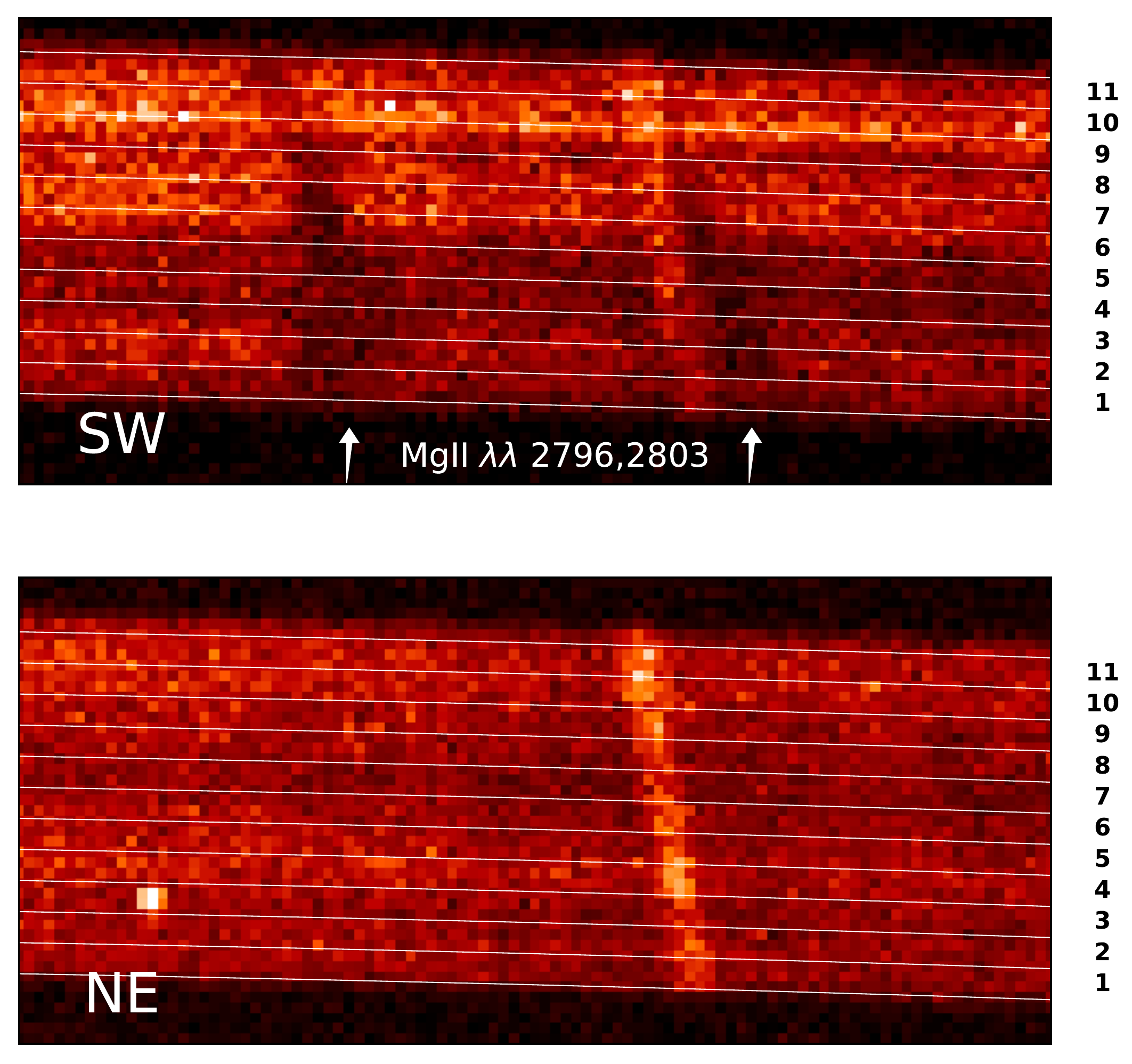} 
\caption{Raw MagE 2D spectra obtained through the SW (upper panel) and NE (bottom panel) slits. 
Each exposure is
  $3\,600$\,s long. 
  Wavelength increases to the right and each spectral pixel
  corresponds to $\approx 22$\,\kms. 
Both spectra are centered at
  $\lambda\approx 4850$\,\AA, the expected position of 
  \mgii$\lambda\lambda$2796,2803 at $z=\zabs$ (indicated by the arrows in the upper panel).  
   \mgii\  absorption is clearly seen all along the SW slit, but not in the NE slit. 
    Moreover, the velocity shift and kinematical complexity of the \mgii\ absorption 
    seems to be a function of the spatial position with respect to G1, which is  located around  SW position \# 2 (see also Fig.~\ref{fig_slits}).  
    The grid tracing the echelle orders corresponds to the eleven spatial  
  positions (pseudo-spaxels) described in the text, with numbers (indicated on the right margin)  increasing from North to South. Each position is
  $0.9\arcsec$ along the slit, and the slit width used was $1.0\arcsec$.   
       A sky line at $4861.32$\,\AA\ blocks
  partially the $2803$\,\AA\ transition, unfortunately, but it otherwise aids the eye to follow the spatial direction on the CCD.
\label{fig_2D}}  
\end{figure}

The data were reduced using a custom pipeline (see details in
\S~\ref{sec:mage_reduction}).  The spectra cover the wavelength range
$3\,300$--$9\,250$ \AA\ at a resolving power $R=4\,500$.  For each slit, 11
calibrated spectra were generated using a $3$-pixel spatial binning,
corresponding to $0\farcs9$ on the sky (see Fig.~\ref{fig_2D}). Such binning
oversamples the seeing, making the spectra spatially independent.  These
spectra define 11 `pseudo-spaxels' in each slit. The spectra were recorded
into three data-cubes of a rectangular shape of $1 \times 11$ `spaxels' of
$1\farcs0\times0\farcs9$ each. Throughout the paper, we use the convention
that the northernmost spaxel in a given slit is its `position 1' (e.g., SW
\#1) and position numbers increase toward the South in consecutive order (see
Figs.~\ref{fig_FOV} to~\ref{fig_2D}).

\begin{figure}
\includegraphics[width=\columnwidth]{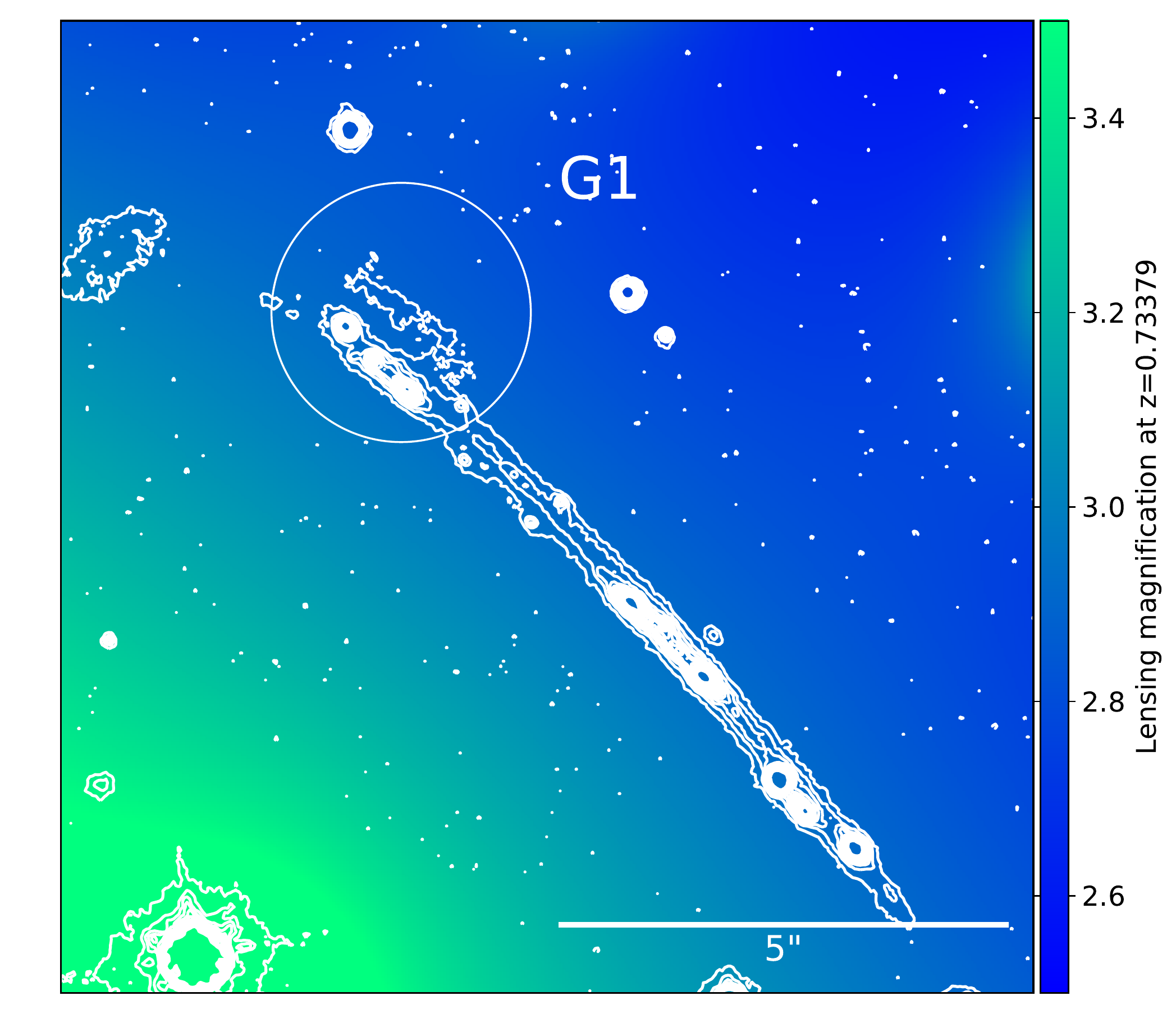}
\caption{Magnification map at $z=\zabs$ (displayed in the image plane). The contours
correspond to the {\it HST} F814W image. We caution that this figure does not show the magnification of the giant arc itself, which is at  a different source redshift.
\label{fig_magnification}}  
\end{figure}

\section{Lens model and absorber-plane geometry}
\label{sec_lens_model}

In this section we describe the lens model used to reconstruct the absorber plane and to  properly define impact parameters.

\subsection{Lens model}
\label{sec:lensing}

The lens model is computed using the public software
\lenstool~\citep{Jullo2007}.  Our model includes cluster-scale, group-scale,
and galaxy-scale halos. The positions, ellipticities, and position angles of
galaxy-scale halos are fixed to the observed properties of the cluster-member
galaxies, which are selected from a color-magnitude diagram using the red
sequence technique~\citep{Gladders2000}. The other parameters are determined
through scaling relations, with the exception of the brightest cluster galaxy
that is not assumed to follow the same scaling. Some parameters of galaxies
that are near lensed sources are left free to increase the model flexibility.
The parameters of the cluster and group scale halos are set as free
parameters. The model used in this work solves for six distinct halos, and
overall uses 100 halos.

We constrain the lens model with positions and spectroscopic redshifts of
multiple images of lensed background sources, selected from our \textit{HST}
imaging and 
lensing analysis in this field will be presented in Sharon et al., in prep.

From the resulting model of the mass distribution of the foreground lens, we derive the lensing magnification and deflection maps that are used in this work.  The deflection map $\vec{\alpha}$  is used to ray-trace the observed positions to a background (source) plane, using the lensing equation: 
\begin{equation}
    \vec{\beta}  = \vec{\theta} - \frac{d_{ls}}{d_s} \vec{\alpha} (\vec{\theta}), \\
\end{equation}
where $\vec{\beta}$ is the position at the background plane, $\vec{\theta}$ is the position in the image plane, and $d_{ls}$ and $d_s$ are  the angular diameter distances from the lens to the source and from the observer to the source, respectively.
In this work, we ray-trace the pixels and spaxels of both the arc and G1, to the absorber plane at $z=\zabs$.

The arc segments are highly magnified and appear at regions close to the
critical curves, where the lensing uncertainties are significant. However, for
the redshift of G1 this region is far enough from the strong lensing regime,
so 
that the lensing potential and its derivatives are smooth (as can be seen in
Fig.~\ref{fig_magnification}) and the uncertainties are reduced.

\subsection{Absorber-plane geometry}
\label{sec_geometry}

\begin{figure}
\includegraphics[width=\columnwidth,clip, trim={2cm 0cm 2cm 0cm},angle=0]{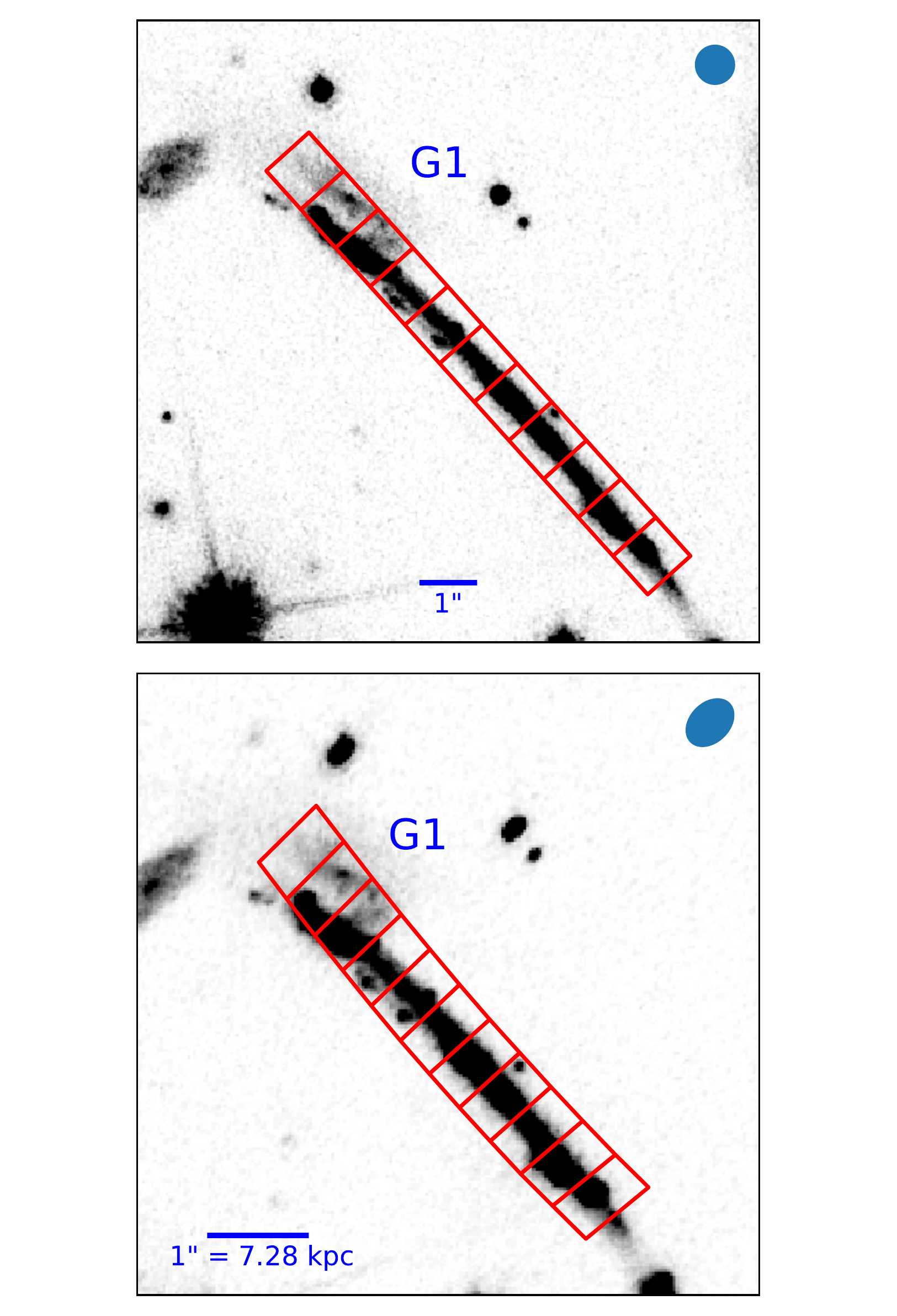}  
\caption{SW slit in the image plane 
  (top) and in the reconstructed absorber plane (bottom).  
The
ground-based observations were taken with a seeing of $0.7^{\prime\prime}$
(indicated by the beam size symbol in the top right of each panel); the background image is that of {\it HST} F814W-band
image that highlights the location and morphology of G1 on both panels. 
In the absorber plane both the F814W image and the slit have
been de-lensed to $z=\zabs$ (see \S~\ref{sec_lens_model}; including the shape of the
PSF, which is used in~\S\ref{oii_emission} to run the galaxy emission model). In this
plane the separation between contiguous MagE spaxels is, on average, $3.2$\,kpc. 
\label{fig_slits}}  
\end{figure}

Fig.~\ref{fig_slits} shows a zoom-in region of the field around G1 in the
image plane (top panel) and in the reconstructed absorber plane at $z=\zabs$
(bottom panel).  For clarity, only the SW spaxels are shown. In the absorber
plane, each spaxel is $\approx 3\times6$~kpc$^2$ in size.

\begin{figure}
\includegraphics[width=\columnwidth,angle=0]{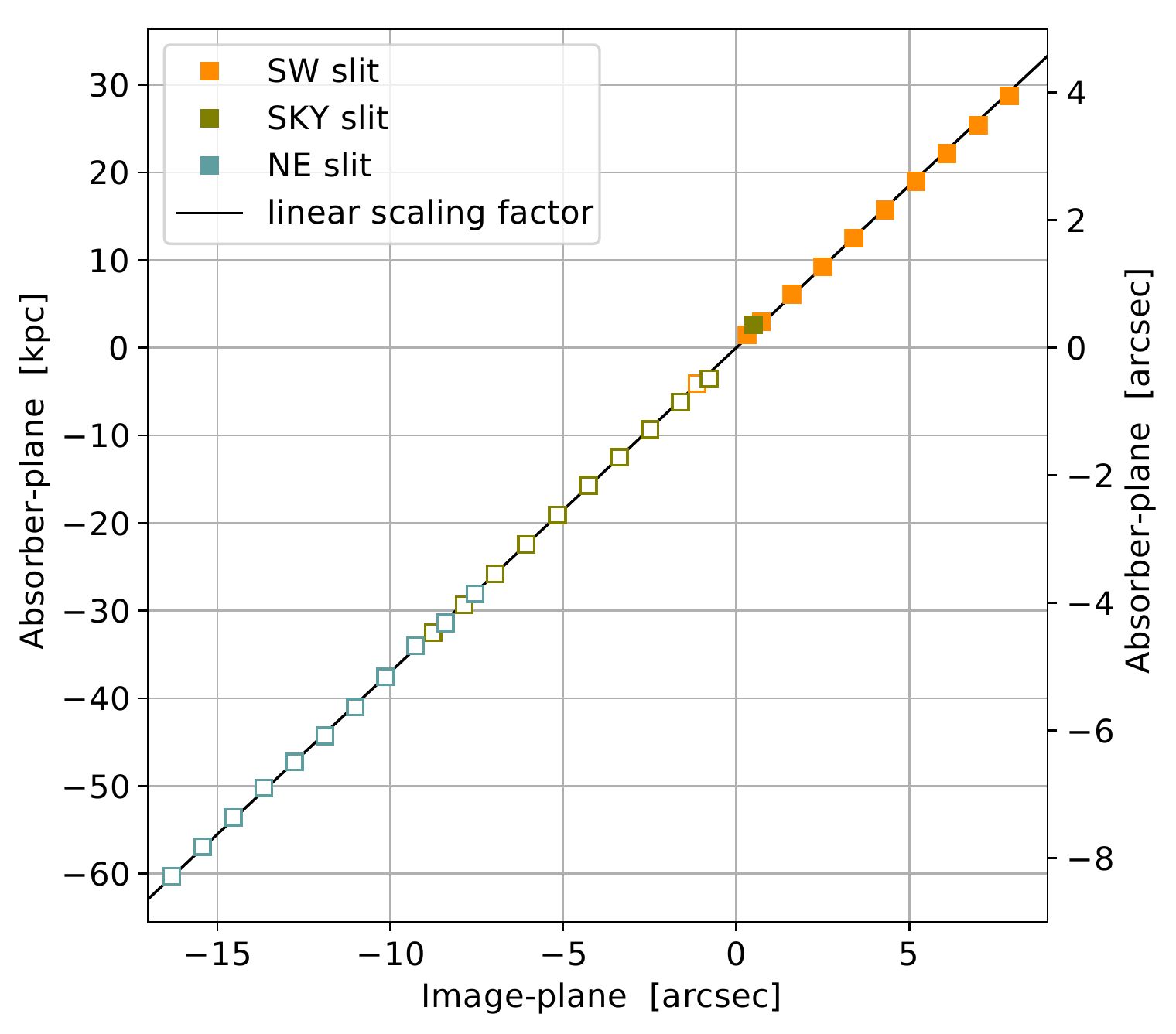} 
\caption{Impact parameters to G1, probed by the MagE spaxels in the image plane
(horizontal scale) and in  the absorber plane (vertical scales). Positions to the
North-East of the G1 semi-minor axis are assigned arbitrarily with negative values and are shown with open symbols. Note that
the transformation from the image to the absorber plane 
is well approximated by 
a constant scale factor (the straight line in the figure). To convert
angular distances into physical distances in the absorber plane  a scale of 
 7.28\,kpc/$\arcsec$ was used.
\label{fig_ip}}  
\end{figure}

Impact parameters, $D$, are defined as the projected distance between the
center of a spaxel and the center of G1.  Impact parameters in arc-seconds are
defined in the reconstructed image.  They are then converted to physical
distances by using the cosmological scale at $z=\zabs$ ($1 \arcsec =
7.28$\,kpc).  For the sake of clarity, we arbitrarily assign negative or
positive values depending on whether the spaxel is to the North-East or to the
South-West of G1's minor axis, respectively. Due to the particular alignment
of galaxy and arc segments, the conversion between impact parameters in the
image and in the reconstructed planes is almost linear (Fig.~\ref{fig_ip}).

Our definition of impact parameter carries three sources of uncertainty.  The
first one comes from the lens model systematics and cosmology; we estimate
this error to be $\approx 5$\%, and therefore to dominate at large impact
parameters.  A second source of error comes from the astrometry, which
introduces an error that dominates at low impact parameters. For instance,
spaxel SW \#1 in Fig.~\ref{fig_slits} does not apparently match any arc signal
in the {\it HST} image.  However, we do measure flux on that spaxel
(Fig.~\ref{fig_2D}), which we render independent from SW \#2, judging from the
different absorption kinematics (Fig.~\ref{fig_stack_mage}). The astrometry is
further discussed in~Appendix~\ref{sect_astrometry}.  These two can be
considered {\it measurement} errors associated with our particular definition
of impact parameter.

A third source of uncertainty comes from the extended nature of the background
source, which is relevant for comparisons with the well defined `pencil-beam'
quasar sight-lines.  Our absorbing signal results from a light-weighted
profile, which in turn is modulated by both the source deflection and the lens
magnification. Thus, our experimental setup faces an inherent source of
systematic uncertainty in the impact parameters (suffered by any observations
using extended background sources).

To account for the last two uncertainties we arbitrarily assign a systematic
error on $D$ of half the spaxel size {\it along the slit}, i.e.,
$\approx1.5$\,kpc in the absorber plane.

\section{Emission properties of G1 at $z=\zabs$}\label{sec:G1}

\begin{table}
\caption{G1 properties}\label{table_G1}
\centering
\resizebox{\columnwidth}{!}{%
\begin{tabular}{lr}
\hline
\hline
\multicolumn{2}{c}{\it From \oii\ emission  and broad-band imaging (see \S~\ref{sec:G1})}\\
Redshift & $z_{\rm abs}=\zabs$ \\
Inclination angle (stars)$^a$ & $i_*=45\pm 5\,\degree$  \\
Position angle (stars)$^a$ & PA$_*=55\pm 3\,\degree$  \\
$B$-band absolute magnitude &  $M_B=-20.49\pm 0.20$\\
$B$-band luminosity$^b$&$L_B=0.14\pm 0.03~L^*_B$ \\
\oii\ flux$^b$ & $f_{\rm OII}= 2.1\times10^{-17}$\,erg\,s$^{-1}$\,cm$^{-2}$  \\
Star-formation rate$^{b,c}$&SFR=$1.1\pm 0.3$\,\msun\,yr$^{-1}$\\
Specific SFR$^{b, c}$ &sSFR=$2.3\pm 0.8\times10^{-10}$\,yr$^{-1}$\\
SF-efficiency$^{b, c, d}$&SFE=$3.5\pm 1.2\times10^{-10}$\,yr$^{-1}$\\
Stellar mass$^b$ & $\log(M_*/\msunm)=9.7\pm 0.3$\\
Halo mass (from M$_*$)$^b$ &  $\log(M_h/\msunm)=11.7\pm 0.3$\\
Virial radius (from M$_h$)$^b$ & $R_{\rm vir}=135$\,kpc\\
&\\
\multicolumn{2}{c}{\it From morphokinematical analysis of \oii\ (see Appendix~\ref{sec:galpak})}\\
Inclination angle (gas)$^a$ & $i_{\rm gas}=49 \pm 3\,\degree$\\
Position angle (gas)$^a$ & PA$_{\rm gas}=70\pm 3\,\degree$\\
Turnover radius (gas)$^{a,e}$ & $r_{t}=3.0 \pm 0.5$\,kpc  \\
Maximum velocity (gas)$^e$ & $v_{\rm max} = 196 \pm 17$\,\kms\\
Velocity dispersion (gas) & $\sigma_v= 9 \pm 4$\,\kms\\
Halo mass (from dynamics) & $\log(M^{\rm dyn}_h/\msunm) = 12.2 \pm 0.1$ \\
Virial radius (from dynamics) & $R_{\rm vir}^{\rm dyn} = 190 \pm 17$\,kpc\\
\hline
\end{tabular}
}
\\
\vspace{1ex}
\begin{minipage}{0.8\columnwidth}
\footnotesize{
\raggedright {\bf Notes:}\\
$^a$ In the reconstructed absorber plane. \\
$^b$ De-magnified quantity using $\mu=2.9$ (see \S~\ref{sec_lens_model}).\\
$^c$ Obscured.\\
$^d$ Using neutral gas mass $\log(M_{\rm HI}/M_{\astrosun})=9.5$ (see \S~\ref{sec:dlas}).\\
$^e$ Defined from the arctan rotation curve: $v(r)=v_{\rm max}\arctan(r/r_t)$.
}
\vspace{2ex}
\end{minipage}
\end{table}

\begin{table*}
\caption{\oii\ emission-line properties in the MagE data}
\centering
\begin{tabular}{ccccc}
\hline
\hline
Slit pos. & $D$     
&  Flux(3729) &  $v$     &  $\Delta v_{\rm FWHM}$   \\
 & (kpc) & (10$^{-20}$erg~s$^{-1}$cm$^{-2}$) & (\kms)            & (\kms)         \\
 (1) &(2) &(3) &(4)& (5)\\
\hline
SW \#1&-3.6&10.6$\pm$ 0.7 &-31.4$\pm$6.1 &169.8\\
SW \#2 & 1.4& 3.3$\pm$ 0.1 & 10.0$\pm$3.6 &171.0\\
SW \#3 & 3.6& 2.5$\pm$ 0.2 & 99.9$\pm$4.1 &153.3\\
SW \#4 & 6.8& 0.8$\pm$ 0.2 &104.3$\pm$ 9.7& 97.5\\
\hline
\end{tabular}\\
\vspace{1ex}
\begin{minipage}{0.53\textwidth}
\raggedright {\bf Notes:}
(1) MagE spaxel position number in the SW slit (see Fig.~\ref{fig_FOV});
(2) Projected physical separation between the center of the MagE spaxel and G1 in the absorber plane;
(3) Total \oii\,$\lambda$3729\,\AA\ flux;
(4) Rest-frame velocity of the \oii\ emission with respect to the systemic redshift, $z_{\rm abs}=\zabs$;
(5) Velocity spread of the \oii\ emission.
\vspace{2ex}
\end{minipage}
\label{table_emission}
\end{table*}

We use the {\it HST} and MUSE datasets to characterize G1. In the following subsections we present the details of these analyses, and Table~\ref{table_G1} summarizes G1's inferred properties.

\subsection{Geometry and environment}

From the source plane reconstruction of the  {\it HST} image (see bottom panel of Fig.~\ref{fig_slits}), 
G1 is a spiral galaxy with well defined spiral arms.  The position angle of the major axis is  PA$=55\degree$ N to E. The
axial ratio is about 0.7, which implies an  inclination angle of
$i=45\degree$.

G1 seems to have no companions nearby.  
We have run an automatic search for emission line sources and found no other
galaxy at this redshift in the MUSE field.  According to our lens model, G1 is
magnified by a factor of $\mu \approx 2.9$. The model does not identify regions
with much lower magnification around G1 (Fig.~\ref{fig_magnification})
implying that no other non-magnified galaxies have been missed by our
automatic search, 
down to a $1\sigma$ surface brightness limit  of $\approx
5\times10^{-19}$ erg~s$^{-1}$~cm$^{-2}$~arcsec$^{-2}$.

\begin{figure}
\includegraphics[width=\columnwidth,angle=0]{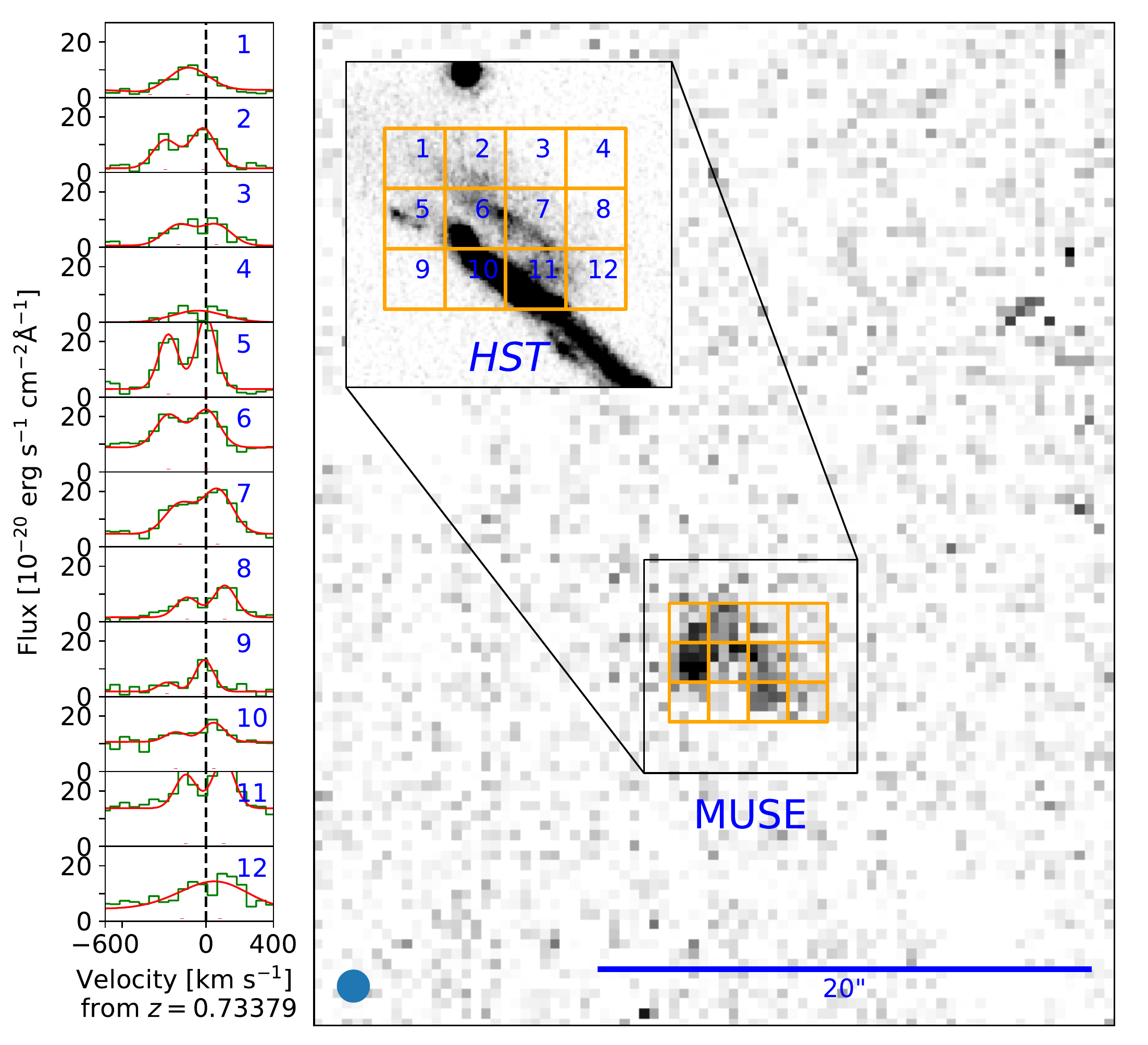}
\caption{{\it Right panel:} \oii\ nebular emission around G1  in the MUSE cube. Stars and foreground objects have been removed. The inset shows G1's stellar emission  as seen in the {\it HST} F814W band. Both images are displayed in the image plane. Yellow boxes are $0\farcs8$ on each side, corresponding to $4\times 4$ MUSE spaxels. The blue circle indicates the seeing FWHM. {\it Left panels: } Gaussian fits to  \oii$\lambda\lambda3727,3729$ at each of the 12 selected regions indicated by the numbered boxes.}
\label{fig_oii}  
\end{figure}

\subsection{{\it HST} photometry}
\label{sec_photometry}

G1 is located in projection close to the bright arc (see Fig \ref{fig_slits});
thus, the photometry is expected to be contaminated. To measure the galaxy
flux we use two different techniques. We first apply a symmetrization approach
in which we rotate the galaxy image, subtract it from the original and clip
any $2\sigma$ positive deviations; this image is finally subtracted from the
original and in this fashion the unrelated emission is
eliminated~\citep{Schade1995}.  The second approach is to obtain the flux from
a masked image that excludes the arc.  From both methods we obtain an average
$m_{F814W} = 21.76 \pm 0.20$ and $m_{F160W} = 22.04 \pm 0.17$, corrected for
Galactic extinction of E(B-V)=0.094\,mag using dust maps of
\citet{Schlegel1998}. The absolute magnitude is computed from the F814W band
which is close to rest-frame B-band, and the small offset is corrected using a
local SBc galaxy template~\citep{Coleman1980}. The absolute magnitude is
$M_B=-20.49$.  Using the luminosity function from DEEP2~\citep{Willmer2006} we
obtain a de-magnified luminosity of $L/L^*_B=0.14$.

Using a standard SED fitting code~\citep{Moustakas2017} we constrain the
(de-magnified) median stellar mass to be M$_*= 4.8 \times 10^{9}$
M$_{\astrosun}$. Using the stellar-to-halo mass relation in~\citep{Moster2010}
we infer a halo mass of M$_h=4.8 \times 10^{11}$ M$_{\astrosun}$, which
corresponds to a virial radius of $R_{\rm vir}\approx135$\,kpc.

\subsection{[O~II] emission}
\label{oii_emission}

Fig.~\ref{fig_oii} shows the nebular \oii\ emission around G1 as obtained from
the MUSE datacube (i.e., in the image plane), from which we define the
systemic redshift.  We fit the \oii$\lambda 3727,3729$ doublet with double
Gaussians in 19 $4\times4$ binned spaxels (of which the brightest 12 are shown
in Fig.~\ref{fig_oii}) and obtain a total (de-magnified) \oii\ flux of $f_{\rm
  OII}=2.1\times10^{-17}$ erg~s$^{-1}$~cm$^{-2}$.  Considering the luminosity
distance to $z=\zabs$ we infer a (obscured) star-formation
rate~\citep{Kennicut1998} of SFR$=1.1$\,\msun\,yr$^{-1}$.  Considering its
redshift and specific star formation, G1 represents a star-forming
galaxy~\citep{Lang2014,Oliva-Altamirano2014,Matthee2019}.

To compare \oii\ emission with \mgii\ {\it absorption} velocities, we map the
MUSE spaxels into the MagE spaxels. In this fashion we make sure we are
sampling roughly the same volumes both in emission and absorption (although
for the reasons outlined in~\S\ref{sec_lens_model} the physical regions are
not constrained within a spaxel, and therefore we cannot establish whether
\oii\ and \mgii\ occur in {\it exactly} the same volumes). We set $v=0$
\kms\ at $z=0.73379$.  The re-mapped cube shows significant \oii\ emission in
MagE spaxels SW \#1 through \#4 (Fig.~\ref{fig_stack_mage}). The fit results
are listed in Table~\ref{table_emission}.

We also perform a morpho-kinematical analysis of G1's \oii\ emission using the
\galpak\ software~\citep{Bouche2015}. The input is a reconstructed version of
the MUSE cube in the absorber plane (see Appendix~\ref{sec:galpak} for
details).  From the model we obtain an independent assessment on the geometry
and halo mass of the galaxy (see Table~\ref{table_G1}). We find a total halo
mass that is somewhat larger than that obtained from the SED fitting, but
consistent within uncertainties. We also find consistency for G1's
inclination. However, the inferred PAs of the major axis differ by $\sim
15\degree$, which should not be a surprise if gas and stars have somewhat
different geometries. We come back to the \galpak\ model
in~\S~\ref{sec_velocities} when we assess the kinematics of the absorbing gas.

\section{Absorption properties of G1 at $z=\zabs$}\label{sec:abs}

This section encompasses the core of the present study. We analyze the
absorption-line properties of G1 according to both absorption strengths and
kinematics in the MagE data.  
We emphasize that MagE blue coverage and resolving power should lead to   
robust equivalent width ($W_0$) and redshift  measurements.

\subsection{MagE absorption profiles}

\mgii\ is
detected in all 11 SW positions and in 2 of the SKY positions. All but 3 (4)
of these detections have also \feii\ (\mgi) detections. In the NE arc-segment, 
we find no absorption in any of the 11 positions down to 
sensitive limits.

\begin{figure}
\includegraphics[width=1.\columnwidth,angle=0]{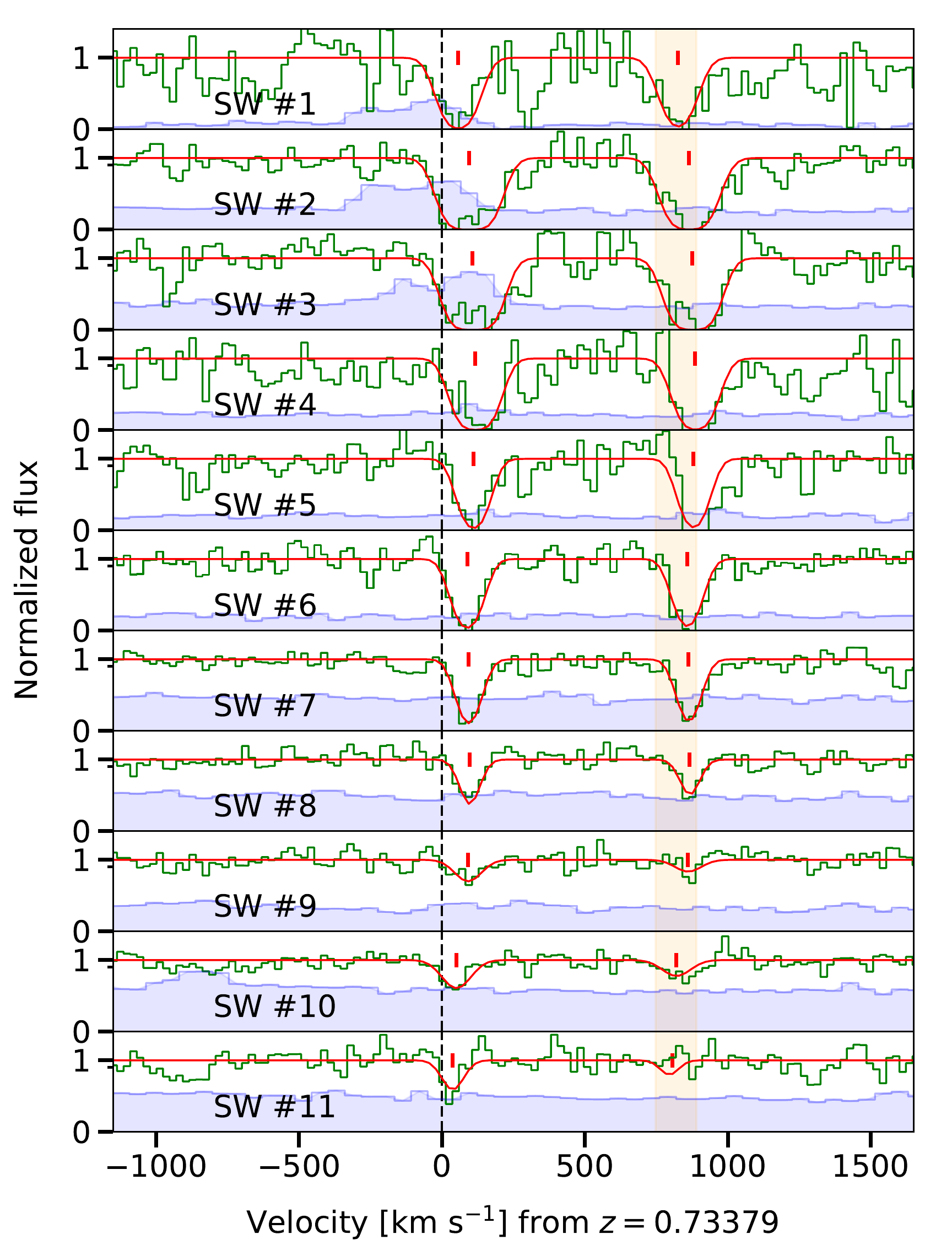} 
\caption{\mgii\ detections in the SW slit. Position numbers of the MagE spaxels are indicated,
  with numbers increasing to the South-West. The center of G1 lies close to SW \#2 (see Figs.~\ref{fig_FOV} and \ref{fig_slits}). The blue shaded spectrum
  corresponds to \oii\ coverage (scaled to fit in the y-axis) as measured with
  MUSE over the MagE spaxels. Only the
  four MagE spaxels that lie 
  closest to G1 (SW positions \#1 to \#4) show noticeable \oii\ (see Table~\ref{table_emission}). The yellow shaded region 
  indicates the position of a sky emission line at $4861.32$\,\AA\ (see Fig.~\ref{fig_2D}).
\label{fig_stack_mage}}  
\end{figure}

To obtain $W_0$ and redshifts, we fit single-component Voigt profiles in each
continuum-normalized spectrum.  The spectral resolution of MagE is not high
enough to resolve individual velocity components and therefore the fits are
not unique; however, using Voigt profiles (instead of Gaussian profiles)
allows us to obtain equivalent widths and accurate velocities via simultaneous
fitting of multiple transition lines. We use the VPFIT package
\citep{Carswell2014} to fit the following lines:
\mgii~$\lambda\lambda2796,2803$, \mgi~$\lambda2852$, and
\feii~$\lambda\lambda\lambda\lambda2600,2585,2382,2374$.  \feii~$\lambda 2344$
was excluded from the analysis because it is in the source's Ly$\alpha$
forest. Possible \caii\ lines are heavily blended with sky lines in the red
part of the spectrum and were not considered either.  In each fit, redshift,
column densities ($N$) and Doppler parameters ($b$) were left free to vary
while keeping all transitions tied to a common redshift and Doppler parameter,
and the same species to a common column density.  We calculate equivalent
widths and their errors from the fitted $N$ and $b$ values using the
approximation provided in \citet[][]{Draine2011}. $W_0$ upper limits for
non-detections are obtained using the formula $W_0(2\sigma) = 2\times{\rm
  FWHM}/\langle S/N \rangle/(1+z)$, where $\langle S/N\rangle$ is the average
signal-to-noise per pixel at the position of the expected line.  The full
velocity spread of the system, $\Delta v_{\rm FWHM}$, is estimated from the
deconvolved synthetic profile of \mgii\ $\lambda 2796$.

The complete set of synthetic profiles and non-absorbed spectral regions is
shown in the Appendix. The fitted parameters are presented in
Table~\ref{table_abslines}.  Aided by the fitted profiles, we do not see
evidence of anomalous multiplet ratios, and therefore assume no partial
covering effects~\citep[e.g.,][]{Ganguly1999,Bergeron2017}.

In Fig.~\ref{fig_stack_mage} we present the \mgii\ absorption profiles and
their fits in the SW slit (the fits are constrained by the \feii\ lines as
well, not shown here but in the Appendix). The fitted profiles feature a clear
transition from stronger (kinematically more complex) to weaker (simpler)
systems, as one probes outwards of G1, i.e., with increasing position number
along the slit.  The errors in velocity, just a few \kms, are small enough to
also reveal a clear shift in the centroid velocities (red tick-marks in the
Figure) that change with position in a non-random fashion. We come back to
these kinematical aspects in~\S\ref{sec_velocities}
and~\S\ref{sec_kinematics}.

\begin{figure}
\includegraphics[width=\columnwidth,angle=0,clip]{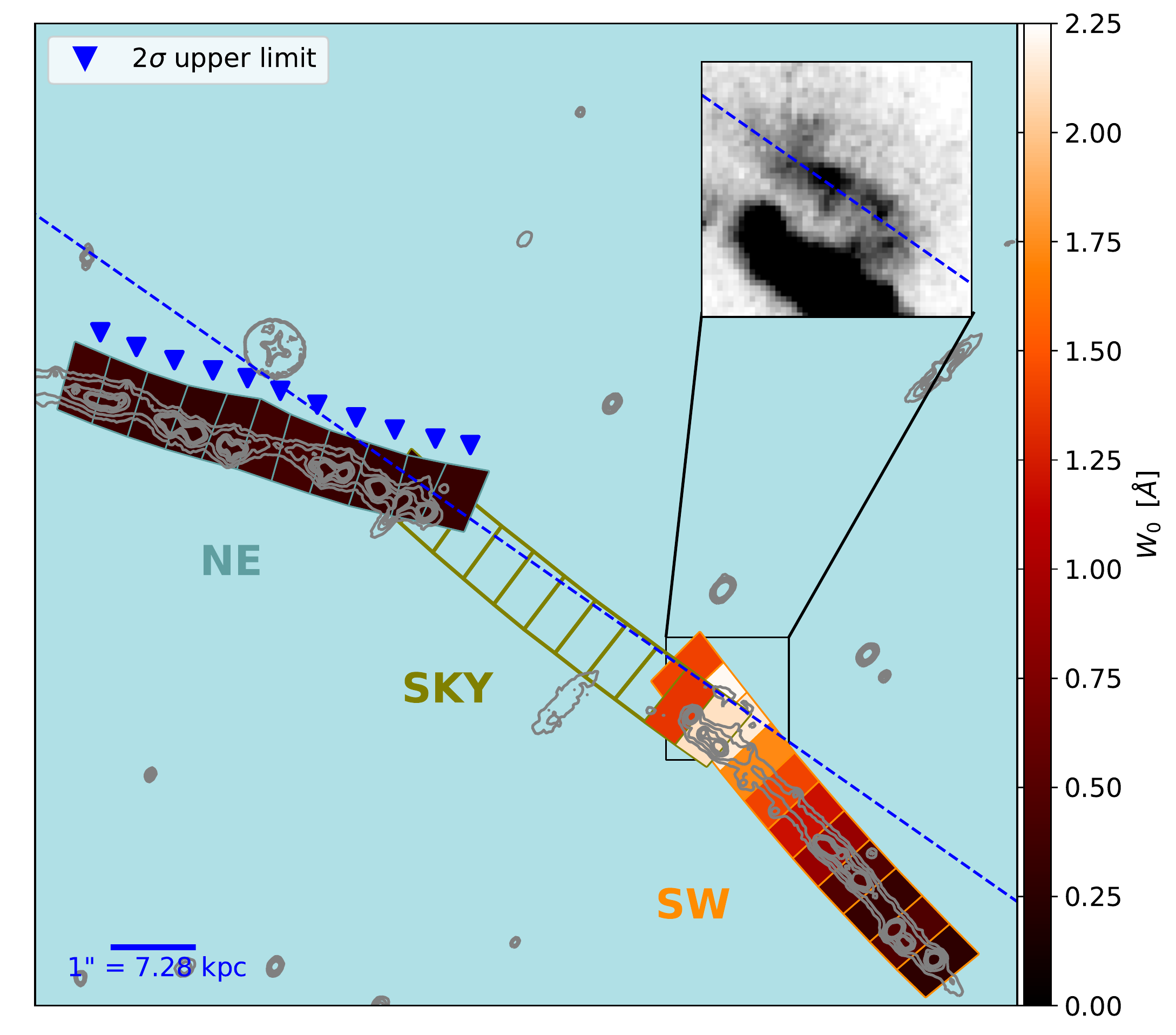} 
\caption{\mgii\ $W_0(2796)$ map in the absorber (de-lensed)
  plane. Each spaxel is $3\times 6$ kpc$^2$. SKY slit positions with no source illumination are shown transparent. Upper limits ($2\sigma$) in the NE slit are indicated with blue triangles).
  The dashed line indicates the projection of G1's semi-major axis at PA=$55\degree$ N
  to E. The inset shows an image from the {\it HST} F814W band in the absorber
  plane. 
\label{fig_rEW_map}}  
\end{figure}

Fig.~\ref{fig_rEW_map} shows the corresponding  map of $W_0(2796)$ in the
(re-constructed) absorber plane. The color of each spaxel is tied to the
rest-frame 
equivalent width when \mgii\ is detected. The blue arrows indicate $2\sigma$
upper limits while the  dashed line indicates G1's position angle.  
This map provides an overall picture of the present scenario: coherent
absorption in a highly inclined disk along its major axis toward the
South-West  direction,
with two detections in the North-East side of G1. Conversely, the NE slit, further
away from G1, shows no detections.

In the following analysis we consider separately the equivalent widths and the
velocities, both as a function of $D$. 

\begin{table*}
\caption{Absorption-line properties in the MagE data.
}\label{table_abslines}
\centering
\begin{tabular}{lcccccc}
\hline
\hline
 Slit pos. & $D$     &  $v$     &  $\Delta v_{\rm FWHM}$ &$W_0(2796)^a$ &  $W_0(2600)^a$ &  $W_0(2852)^a$  \\
        & (kpc) & (\kms) & (\kms) &(\AA)        & (\AA)        & (\AA) \\
(1) & (2) & (3) & (4) & (5) & (6) & (7) \\
\hline
SW \#1&-3.6&56.7$\pm$4.1&155.6&1.55$\pm$0.18&1.62$\pm$0.06&1.7$\pm$0.2\\
SW \#2&1.4&95.4$\pm$2.9&230.7&2.27$\pm$0.15&1.97$\pm$0.04&1.01$\pm$0.11\\
SW \#3&3.6&107.2$\pm$4.3&221.4&2.19$\pm$0.27&1.63$\pm$0.06&0.64$\pm$0.1\\
SW \#4&6.8&116.5$\pm$5.2&179.2&1.79$\pm$0.27&1.57$\pm$0.07&0.56$\pm$0.13\\
SW \#5&9.9&110.3$\pm$3.8&103.8&1.23$\pm$0.41&0.98$\pm$0.12&0.25$\pm$0.08\\
SW \#6&13.2&88.4$\pm$2.8&110.8&1.19$\pm$0.15&0.74$\pm$0.02&0.15$\pm$0.06\\
SW \#7&16.5&93.4$\pm$2.4&121.2&0.91$\pm$0.25&0.44$\pm$0.03&0.13$\pm$0.04\\
SW \#8&19.8&97.0$\pm$5.0&53.6&0.52$\pm$0.13&0.21$\pm$0.03&$<$0.14\\
SW \#9&23.0&91.1$\pm$12.6&82.9&0.33$\pm$0.05&$<$0.11&$<$0.19\\
SW \#10&26.3&50.5$\pm$6.7&91.7&0.47$\pm$0.04&$<$0.12&$<$0.1\\
SW \#11&29.7&37.5$\pm$9.0&50.3&0.37$\pm$0.03&$<$0.21&$<$0.15\\

SKY \#1&-32.5&...&...&$<$0.28&$<$0.56&$<$0.3\\
SKY \#10&-3.8&60.5$\pm$5.7&168.5&1.35$\pm$0.12&2.02$\pm$0.06&1.08$\pm$0.15\\
SKY \#11&3.3&94.2$\pm$4.8&159.9&2.14$\pm$0.19&2.24$\pm$0.09&1.33$\pm$0.11\\

NE \#1&-61.5&...&...&$<$0.22&$<$0.3&$<$0.26\\
NE \#2&-58.0&...&...&$<$0.17&$<$0.29&$<$0.25\\
NE \#3&-54.6&...&...&$<$0.14&$<$0.2&$<$0.23\\
NE \#4&-51.2&...&...&$<$0.12&$<$0.21&$<$0.21\\
NE \#5&-48.1&...&...&$<$0.15&$<$0.24&$<$0.23\\
NE \#6&-45.2&...&...&$<$0.21&$<$0.42&$<$0.33\\
NE \#7&-41.9&...&...&$<$0.23&$<$0.3&$<$0.39\\
NE \#8&-38.5&...&...&$<$0.16&$<$0.23&$<$0.21\\
NE \#9&-34.8&...&...&$<$0.13&$<$0.2&$<$0.18\\
NE \#10&-32.0&...&...&$<$0.12&$<$0.2&$<$0.16\\
NE \#11&-28.8&...&...&$<$0.16&$<$0.25&$<$0.21\\

\hline
\end{tabular}\\
\vspace{1ex}
\begin{minipage}{0.69\textwidth}
\raggedright {\bf Notes:} 
(1) MagE slit and spaxel position number (see Fig.~\ref{fig_FOV});
(2) Projected physical separation between the center of the MagE spaxel and G1 in the absorber plane; negative values indicate
positions to the North-East of G1's minor axis;
(3) Rest-frame velocity centroid of the absorption with respect to the systemic redshift, $z_{\rm abs}=\zabs$;
(4) Velocity spread of the \mgii~$\lambda 2796$ absorption.
(5) Rest-frame equivalent width of the \mgii~$\lambda 2796$ absorption.
(6) Rest-frame equivalent width of the \feii~$\lambda 2600$ absorption.
(7) Rest-frame equivalent width of the \mgi~$\lambda 2852$ absorption.\\
$^a$ Non-detections are reported as $2\sigma$ upper limits.
\vspace{2ex}
\end{minipage}
\end{table*}

\subsection{Equivalent widths versus impact parameter}

\begin{figure}
\includegraphics[width=\columnwidth,angle=0]{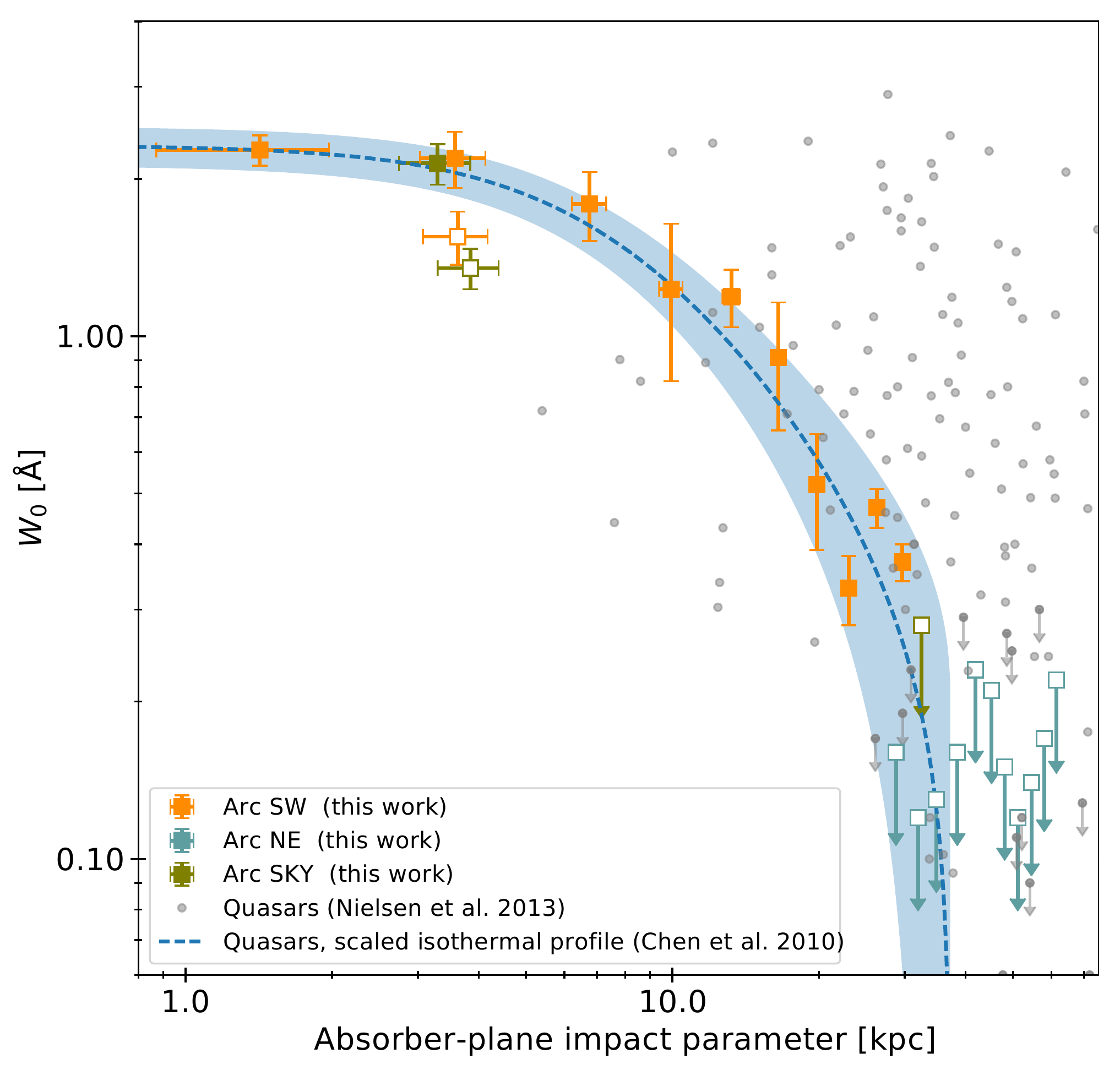}
\caption{\mgii\ $\lambda 2796$ rest-frame equivalent width as a function of impact
  parameter $D$ (in the absorber plane) for SW, NE, and SKY slits. Non-detections
  are reported as $2\sigma$ upper limits. Positions to the North-East of the G1 minor axis are depicted with open
  symbols. 
   Measurement uncertainties in $D$ (\S~\ref{sec_lens_model}) come
  from the astrometry (horizontal error bars) and from the lens model (represented by symbol sizes).
  For comparison with the quasar statistics, data points from~\citet{Nielsen2013} are displayed (grey symbols).
 The dashed curve is a scaled version of the  isothermal density  profile
  from~\citet{Chen2010} using $L=0.14~L^*$ and the  shaded region is the RMS
  of the differences between model and data 
(see \S~\ref{sec_isothermal} for details).
\label{fig_rEW_D}}  
\end{figure}

Fig.~\ref{fig_rEW_D} summarizes the first of our main results. It shows an
anti-correlation between \mgii\ $\lambda 2796$ equivalent width and impact
parameter~\citep[e.g.,][]{Chen2010,Nielsen2013} along the three slit
directions used in this work. Thanks to the serendipitous alignment of G1 and
the arc segments, this is the first time such a relation can be observed in an
individual absorbing galaxy along its major axis.

Noteworthy, there appears to be more coherence toward \pks\ along the SW slit than in the
system studied toward \rcs\ (Paper I), in the sense that all SW positions have
positive detections, having no non-detections down to $\approx 0.2$\,\AA, our
$2\sigma$ detection limit. Since we are probing here (1) along the major axis
of a disk galaxy, and (2) smaller impact parameters, 
the observed coherence probably indicates that the gas in the disk (this arc) is less
clumpy than further away in the halo (\rcs).

We compare these arc data with the statistics  of  quasar absorbers in \S~\ref{sec:discussion}.

\begin{figure*}
\includegraphics[width=1.01\textwidth,clip,angle=0]{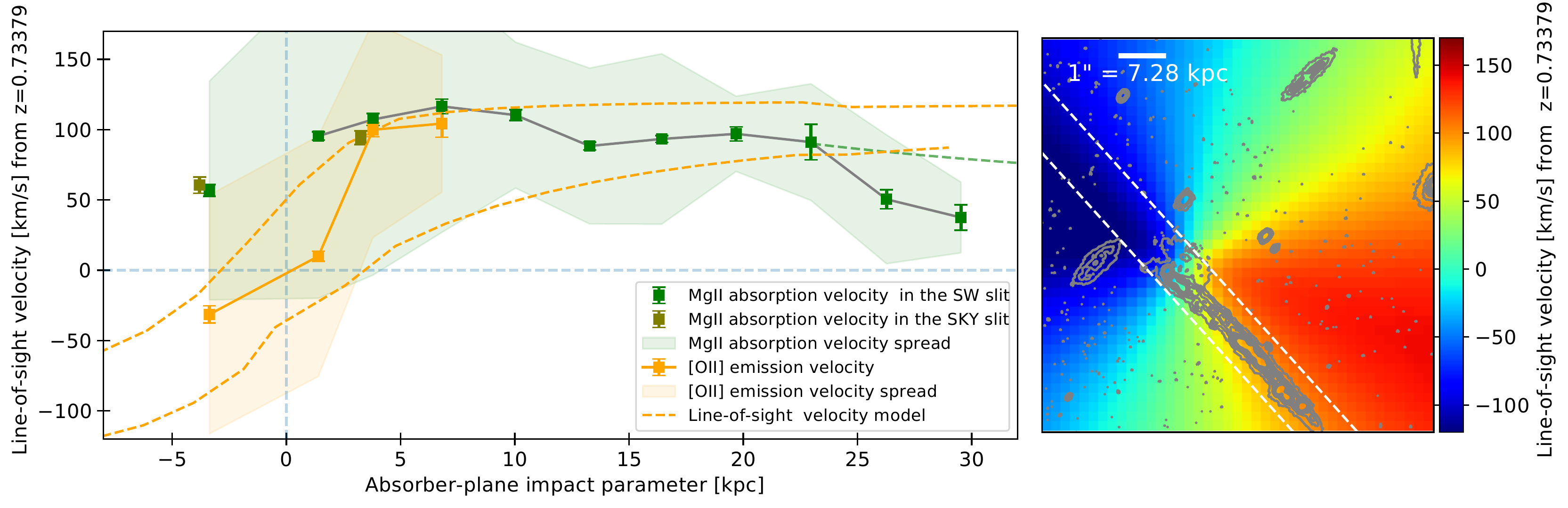}
\caption{{\it Left panel:} Measured line-of-sight velocities versus absorber-plane impact parameter to G1.
Green symbols correspond to \mgii+\feii\ absorption, as
measured in the MagE spectra. The green shaded 
region indicates the projected absorption velocity spread $\Delta v_{\rm FWHM}$ at each
position. The green dashed curve corresponds to Keplerian fall off from the flat part of
the rotation curve. 
Orange symbols correspond to \oii\ emission, as
measured in the MUSE spectra through apertures 
that match SW 
spaxels \#1 to \#4. 
The orange shaded region corresponds to the projected emission velocity spread $\Delta v_{\rm FWHM}$. 
The orange dashed curves are rest-frame line-of-sight velocities drawn from the
\oii\ emission model at the slit edges shown in the right panel. 
Distances to the North-East of G1's semi-minor axis have been 
arbitrarily assign negative values in the impact parameters (see Fig.~\ref{fig_ip}). Impact parameters uncertainties are
the same as in Fig.~\ref{fig_rEW_D}.
{\it Right panel:}
Model line-of-sight velocities in \kms\ from $z=\zabs$ (\S~\ref{oii_emission}).  
The  dashed lines indicate the pseudo-slit used to
extract the velocity limits we display in the left panel. 
The contours correspond to the {\it HST} F814W image in the  reconstructed  absorber plane. 
\label{fig_vel_D}}  
\end{figure*}

\subsection{Gas velocity versus impact parameter}
\label{sec_velocities}

Fig.~\ref{fig_vel_D} displays our second main result. The left panel shows
\mgii-\feii\ absorption velocities in the SW and SKY slits (green and olive
colors, respectively) and \oii\ emission velocities (orange colors) as a
function of impact parameter, $D$.  The emission velocities come from
\oii\ fits in apertures that match SW spaxels \#1 to \#4 (only the four
closest spaxels to G1 show significant \oii; Fig.~\ref{fig_stack_mage}).
Error bars indicate the uncertainty in the velocity centroid, while the shaded
region indicates the projected velocity spread. Note that no spaxel coincides
with $D=0$ kpc.  In this and next figures we treat impact parameters on the NE
side of G1's minor axis as negative quantities (and hence we get rid of the
open symbols). This choice spots apparent rotation around G1, that we discuss
below.  Given the alignment between the arc and the G1's major axis, such a
plot can be considered a rotation curve.  This is the first rotation curve of
absorbing gas measured in such a distant galaxy.

Perhaps the most striking feature in the left panel of Fig.~\ref{fig_vel_D} is
the {\it decline} in velocity at SW spaxels \#10 and \#11.  To explore
possible gas rotation, we use our 3D model of \oii\ emission
(\S~\ref{oii_emission}) and obtain a line-of-sight velocity map at any
position near G1 (right panel of Fig.~\ref{fig_vel_D}). This model might not
be unique, but it does serve our purpose of extending it to larger distances
for comparison with the absorbing gas.  The line-of-sight velocities allowed
by the model within an aperture that matches the SW slit are represented in
the left panel by the dashed curves. It can be seen that most
\mgii\ velocities are well comprised by the model velocities, indicating
co-rotation of the absorbing gas out to $D\approx 23$\,kpc.  The exception are
velocities at SW spaxels \#1 (discussed in \S~\ref{sec_kinematics}), and \#10
and \#11 (\S~\ref{sec_accretion}).

\subsection{Summary of absorption properties}

Before proceeding to the discussion, it is useful to consider an overview of
the observables by including the other two absorption species detected and
their equivalent-width ratios. Such absorption-line summary is shown in
Fig.~\ref{fig_ratios}, where the upper panel is a simpler version of the left
panel in Fig.~\ref{fig_vel_D}, the middle panel joints equivalent widths of
the 3 species studied in this work, and the bottom panel shows $W_0$
equivalent-width ratios.  We concentrate on the standard ratios \rfemg $\equiv
W_0(2600)/W_0(2796)$ and \rmgimg $\equiv W_0(2852)/W_0(2796)$, bearing in mind
that Mg is an $\alpha$ element and therefore chemical enrichment could affect
those ratios.

From the middle panel it can be seen that, like for \mgii, \feii\ and
\mgi\ equivalent-widths also anti-correlate with $D$. This is expected, since
such species have similar ionization potentials and are most likely
co-spatial~\citep{Werk2014}.

From the bottom panel of Fig.~\ref{fig_ratios}, both \rfemg\ and
\rmgimg\ exhibit a general decrease as we probe further out of G1. This is
more evident in \rfemg, which is above $0.5$ out to SW\#6, and below such
threshold beyond. The trend seems real even excluding position SW\#1, which is
the only measurement above unity (see~\S~\ref{sec_kinematics}).  In the
large-distance end, the two outermost positions have comparatively low
\rfemg\ values.

\section{Discussion}\label{sec:discussion}

In this section, we synthesize the various observables of G1's CGM.  The
discussion revolves around what the observed equivalent widths, kinematics,
and equivalent width ratios tell us about the origin of the \mgii-\feii\ gas.
It also highlights the complementarity between our technique and other CGM
probes.

\subsection{Evolutionary context}

G1 seems to be an isolated, sub-luminous ($0.1L^*_B$) star-forming
($>1.1$\,\msun\,yr$^{-1}$) disk-like galaxy.  Fig.~\ref{fig_oii} shows that
the \oii\ emission is confined to the optical surroundings, while
\mgii\ absorption is detected much further out, at least in the direction of
the SW slit.  This suggests that G1 has recently experienced a burst of
star-formation, which is detached from the older (and more ordered) cool gas.
This is analogous to local galaxies, where H$\alpha$ (also a proxy for star
formation) is not necessarily associated with \hi\ (as detected via 21-cm
observations, and here considered to be traced by \mgii), which is usually
more extended~\citep{Bigiel2012, Rao2013}. Therefore, the offset seen toward
\pks\ should not be surprising for a formed disk still experiencing star
bursts, much similar to \mgii-selected galaxies detected in
emission~\citep{Noterdaeme2010,Bouche2007}.

For comparison with the local Universe, our $W_0(2796)$ measurements are
$\approx 3$ times higher than those found in M31 (similar halo mass, similar
inclination, major axis quasar sightlines) by~\citet{Rao2013} at similar
impact parameters.  Such differences might have an evolutionary or
environmental origin, with G1 bearing a larger gaseous content.

\subsection{Spatial structure of the CGM}

\subsubsection{Direct comparison with quasar and galaxy surveys}

The grey points in Fig.~\ref{fig_rEW_D} are drawn from the sample of 
182 quasar absorbers in~\citet{Nielsen2013}. Note that our data provide seven independent 
measurements to the sparsely populated interval $D<10$\,kpc. 

In general, our data falls within the quasar scatter, but that scatter is much
larger than what we see across the arc.  The smaller arc scatter cannot be due
only to our particular experimental design.  Even if the arc data result from
a light-weighted average (over a spaxel area) the spaxels are independent of
each other and therefore cannot falsify spatial smoothness on the scales shown
in Fig.~\ref{fig_rEW_D}.

In Paper I, we found a similar situation toward \rcs.  These cases strongly
suggest that the scatter in $W_0^{\rm quasar}$ is not intrinsic to the CGM but
rather dominated by the heterogeneous halo population, in which gas extent and
smoothness is a function of host-galaxy intrinsic
properties~\citep{Chen2008,Chen2010,Nielsen2013,Nielsen2015,Rubin2018b} and
orientation~\citep{Nielsen2015}.  It should therefore not be a surprise that
quasar-galaxy samples exhibit more scatter than the present case. Furthermore,
the same should be true for other extended probes of the CGM like background
galaxies~\citep{Steidel2010,Bordoloi2011,Rubin2018a,Rubin2018b}, which also
provide single lines-of-sight~\citep[the exception being the handful of cases
  where background galaxies resolve foreground halos;
][]{Diamond-Stanic2016,Peroux2018}.

\subsubsection{Isothermal-profile model}
\label{sec_isothermal}

We also compare our data with a physically-motivated model.  The dashed line
in Fig.~\ref{fig_rEW_D} shows a 4-parameter isothermal profile with finite
extent, $R_{\rm gas}$, developed by~\citet{Tinker2008} to describe $W_0^{\rm
  quasar}(D)$.  The isothermal profile was first motivated to model the
observed distribution of dynamical mass within $\approx 30$\,kpc of nearby
galaxies~\citep{Burkert1995}.  \citet{Chen2010} fitted such a profile to a
sample of 47 galaxy-\mgii\ pairs and 24 galaxies showing no \mgii\ absorption
at $10<D<120 h^{-1}$\,kpc and obtained the scaling relation $R_{\rm gas} =
74\times(L/L_*)^{0.35}$\,kpc.

We test this model on our arc data by imposing the profile to pass through the
$W_0$ value of the closest spaxel to G1 (SW \#2). We use $L/L_*=0.14$ (see
\S~\ref{sec_photometry}) and set the model amplitude to fit $W_0(2796)=
2.27\pm 0.15$\,\AA\ at $D=1.4$\,kpc, leaving the 3 other model parameters
in~\citet{Chen2010} unchanged. The dashed line in Fig.~\ref{fig_rEW_D} shows
that the isothermal model nicely fits our arc data (RMS$=0.19$ \AA); moreover,
it fits the data not only at the closest spaxel (by construction), but also at
almost all impact parameters (excepting the two measurements to the
``opposite'' side of G1; see next subsection).  This is remarkable, since we
are fitting a single halo with an isothermal profile that fits the quasar
statistics at $D>10$\,kpc, extrapolated to smaller impact parameters.

The fit has important consequences for our understanding of gaseous halos.
First, it validates an isothermal gas distribution over the popular
Navarro-Frenk-White~\citep[NFW; ][]{Navarro1997} profile, which does not
predict a flat $W_0$-$D$ relation at small $D$. This is the first time we can
firmly rule out a NFW model for the cool CGM, thanks to our several detections
at $D<10$\,kpc in a single system.  Incidentally, the fit also lends support
to CGM models that adopt a single density profile~\citep[e.g.,][]{Stern2016}.
Secondly, it suggests that G1's CGM is representative of the \mgii-selected
absorber population, since it can be modeled with parameters that result from
quasar absorber averages and over a wide redshift range. And third, it reveals
that the scatter seen in the overall population includes an {\it intrinsic}
component, likely due to CGM structure on scales of tens of kpc.  It seems
timely to verify these fundamental points with more measurements at small-$D$,
including single detections toward unresolved background sources.

\subsubsection{kpc scales}

The overlap of the SKY and SW slits (Fig.~\ref{fig_rEW_map}) helps us to
qualitatively assess variations in $W_0(2796)$ around G1 on kpc
scales. Firstly, SKY positions \#10 and \#11 partially overlap with SW
positions \#1 and \#2, respectively.  The corresponding equivalent widths,
though, show no significant differences (see Fig.~\ref{fig_rEW_D}), suggesting
that close to G1 (within a few kpc) the gas is smooth on scales of $\approx
1$\,kpc, which is roughly the offset between the aforementioned SKY and SW
spaxels.  This could be due to a covering
factor~\citep{Steidel1997,Tripp2005,Chen2008,Kacprzak2008,Stern2016} close to
unity at small impact parameters ($D\lesssim 10$\,kpc).

\subsubsection{Isotropy}
The two measurements to the ``opposite'' side of G1 (i.e., to the North-East
of G1; open symbols in Fig.~\ref{fig_rEW_D}) depart by 2-3$\sigma$ from the
trend shown by the SW positions to the South-West of G1 at the same impact
parameter (noting that the difference is within the typical scatter reported
toward quasar sightlines at larger distances). This indicates that the gas is
not homogeneously distributed around G1, even at these small distances.

We are not able to test isotropy of the \mgii\ gas on scales between
$4<D<29$\,kpc, unfortunately, due to the lack of arc signal right to the
North-East of G1. However, NE position \#11 is located $29.3$ kpc away from
G1, just as far as SW position \#11 on the other side, and yet it shows no
\mgii\ down to a stringent $2\sigma$ limit of $0.16$ \AA\ ($\log N/{\rm
  cm}^{-2}=12.7$), while the SW position has a significant detection at twice
that value. This situation is remarkable, since NE \#11 appears in projection
on top of the major axis (Fig.~\ref{fig_rEW_map}), while SW \#11 lies around
7\,kpc away in projection from the same axis. The NE non-detection comes then
even more unexpected, under the assumption of isotropy. We conclude that the
gas traced by \mgii, to the extent that we can measure it, is either (1) not
isotropically distributed, or (2) distributed in a disk which is not aligned
with the optical disk, or (3) is confined to a (spherical?) volume
$\lesssim30$\,kpc in size along G1 major axis. This latter option implies that
SW\#11 absorption might have an external origin, a possibility we address
below.

\begin{figure}
\includegraphics[width=1.\columnwidth,angle=0]{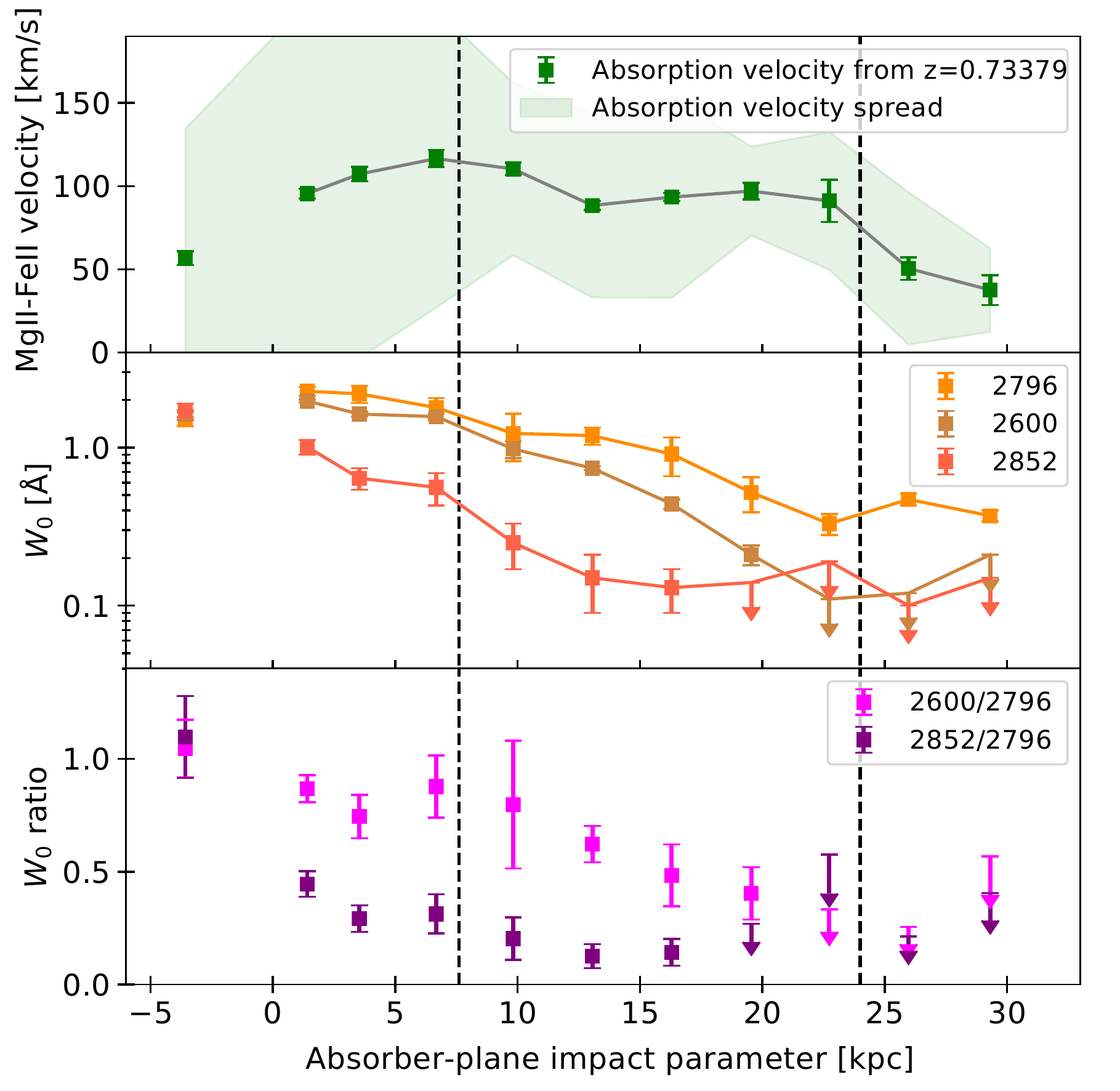}
\caption{Summary of MagE absorption-line properties at $z=0.73379$ toward \pks, as a function of impact
parameter $D$ from G1. Only SW detections are shown. The only impact parameter to
the North-East of G1's minor axis has been flipped the sign. {\it Upper panel:}
Velocity of \mgii+\feii\
line centroids (same as in Fig.~\ref{fig_vel_D}, left panel). {\it Middle panel:}
Rest-frame equivalent width of
\mgii~$\lambda$2796, \feii~$\lambda$2600, and \mgi~$\lambda$2852. {\it
Bottom panel:} Equivalent-width ratios. The vertical dashed lines indicate
the transitions between the absorption regimes proposed in
\S~\ref{sec_kinematics}, i.e., from left to right: disk, disk+inner-halo, and outer
halo absorption.
\label{fig_ratios}}
\end{figure}

\subsection{Kinematics of the absorbing gas}
\label{sec_kinematics}

To the South-West of G1 the absorption signal extends out to $\approx 8$
optical radii along the major axis.  Detecting extraplanar gas at $z=0.7$ has
important consequences for our understanding of disk formation and gas
accretion~\citep[e.g.,][]{Bregman2018,Stewart2011a,Stewart2011}.  The gas
traced by \mgii\ shows clear signs of co-rotation (Fig.~\ref{fig_vel_D}),
suggesting that the shape of the rotation curve is not necessarily governed by
a combination of outflows in less massive halos, as we see here a more ordered
rotating disk. Our data also confirm the rotation scenario unveiled by
simulations~\citep[e.g.,][]{Stewart2011} and also proposed for observations
of disk-selected quasar absorbers at $z\sim 1$~\citep[e.g.,][]{Steidel2002,Ho2017,Zabl2019}.

Based on the line centroids at velocity $v$ (left panel in
Fig.~\ref{fig_vel_D}), and excluding the kinematically detached position SW\#1
(discussed below), we identify three distinct absorption regimes: (1) disk
absorption at $D \lesssim 10$\,kpc, where velocities rise to $\approx
110$\,\kms; (2) disk+inner-halo absorption at $10 \lesssim D\lesssim 20$\,kpc,
where velocities remain flat; and (3) outer-halo absorption at $D\gtrsim
20$\,kpc, where velocities fall down `back' to $v=0$ \kms.

Interestingly enough, the three proposed regimes correlate with the
kinematical complexity of the absorption profiles. In fact, based on the
absorption profiles in Fig.~\ref{fig_stack_mage}, the disk absorption
corresponds to SW positions \#2 to \#4, in which $\Delta v_{\rm
  FWHM}\approx200$\,\kms, suggesting several velocity components (also note
that position \#4 corresponds to the first spaxel beyond the stellar radius;
Fig.~\ref{fig_slits}). Then, the disk+halo absorption corresponds to positions
\#5 to \#9, with somewhat simpler absorption kinematics and smaller $\Delta
v_{\rm FWHM}$ values, suggesting fewer velocity components. We emphasize that
we presently cannot resolve individual velocity components and thus $v$ and
$\Delta v_{\rm FWHM}$ must be considered spectroscopic (and spatial;
see~\S~\ref{sec_geometry}) averages.

The dashed lines in the left panel of Fig.~\ref{fig_vel_D} show that the two
aforementioned regimes are explained, to some extent, by our rotation model.
Conversely, SW positions \#10 and \#11 have the lowest velocity offsets and
spreads, and cannot be explained with rotation, even in the Keplerian
limit (green dashed line in Fig.~\ref{fig_vel_D}). Such `outer-halo' absorption is one of the most striking signature in
the present data, which we discuss in \S~\ref{sec_accretion}.

Finally, SW \#1 also stands out.  This position shows a significantly higher
velocity offset ($\sim 90$\,\kms) than the \oii\ emission, suggesting the
dominant absorbing clouds are not tracking the rotation (the same may be also
true for part of the SW \#2 absorption). The overlapping spaxel SKY \#10 shows
a consistent velocity, meaning that the measurements are robust. Such kind of
offsets are rarely observed in SDSS stacked spectra~\citep{Noterdaeme2010},
suggesting their covering factor is low.  The arc positions also show the
highest \rfemg\ values in our sample, which can be explained if the gas is
more enriched and processed. These two features conspire in favor of a
galactic-scale outflow~\citep{Steidel2010,Kacprzak2012,Shen2012,Fielding2017}
in one of the velocity components, which is escaping G1 in the line-of-sight
direction. Moreover, the spaxels show significant \oii\ flux, and therefore
might be co-spatial with star-forming regions, from which supernova-driven
winds are expected to be launched~\citep[e.g.,][]{Fielding2017,Nelson2019}.

\subsection{Gradient in chemical enrichment?}

Some of the \rfemg\ values in Fig.~\ref{fig_ratios} are exceptionally high
compared with the literature~\citep{Joshi2018,Rodriguez-Hidalgo2012}.  Systems
selected in the SDSS by having \rfemg$>0.5$ are found to probe lower impact
parameters; moreover, there seems to be a distinction between absorbers
associated with high or low SFR depending on whether this ratio is above or
below 0.5, respectively~\citep{Noterdaeme2010,Joshi2018}.  Our particular
experimental setup confirms this trend in the present host galaxy: the four
closest positions to G1 show simultaneously the strongest \oii\ emission
(Fig.~\ref{fig_stack_mage}) and the highest \rfemg\ values (all above 0.5;
Fig.~\ref{fig_ratios}). Furthermore, \rfemg\ seems to show a negative gradient
outwards of G1.

Equivalent widths of saturated lines are known to be a function of the number
of velocity components~\citep{Charlton1998,Churchill2000}, rather than of
column density, $N$.  The present spectra do not allow us to resolve such
clouds nor to get at their $N$-ratios, making it hard to assess unambiguously
the physical origin of the \rfemg\ gradient.  Nevertheless, $N$-ratios must
have an effect on \rfemg. Speculating that both kinematics and line-saturation
affect \mgii$\lambda 2796$ and \feii$\lambda 2600$ similarly at a fixed impact
parameter, a gradient in \rfemg($D$)\ should globally reflect the same trend
in $N($\feii)/$N($\mgii).

$N($\feii)/$N($\mgii) is driven by three factors: (a) ionization: but
assuming \nhi$\gtrsim 19$\,cm$^{-2}$ at $D\lesssim20$\,kpc $\approx
0.1$\,$R_{\rm vir}$~\citep{Werk2014}, ionization is seemingly the less important
factor~\citep{Giavalisco2011,Dey2015}; (b) dust: Mg is less depleted than Fe~\citep{Vladilo2011,DeCia2016};
therefore one expects $N($\feii)/$N($\mgii) (or \rfemg) to {\it
  increase} outwards of G1, which we do not observe; and (c) chemical
enrichment: $\alpha$/Fe decreases as $Z$ increases; therefore,
$N($\feii)/$N($\mgii) (\rfemg) should decrease outwards of G1, which we do observe.

We conclude that we are likely facing the effect of a negative 
gradient in chemical enrichment, with the outermost positions being less
chemically evolved than those more internal to G1. 
Using high-resolution quasar spectra, in a
sample of star-forming galaxies \citet{Zahedy2017} find evidence for a negative gradient in 
 $N($\feii$)/N($\mgii$)$ as well; however, their  ratios
fall down (statistically) at larger distances ($\sim 100$ kpc) than probed here around a single galaxy.
Since \citet{Zahedy2017} galaxy sample is a few to ten times more luminous
than G1, the different scales are likely explained by the  luminosity dependence of $R_{\rm gas}$~\citep[e.g.,][]{Chen2010}.

\subsection{Damped Ly$\alpha$ systems}\label{sec:dlas}

\mgii\ systems having \rfemg$>0.5$ and $W_0(2852)>0.1$
\AA\ have been proposed~\citep{Rao2006,Rao2017} to select damped Ly$\alpha$
systems ~\citep[DLAs; mostly neutral absorption systems having 
$\log$\nhi$>20.3$\,cm$^{-2}$; e.g.,][]{Wolfe2005} at $z<1.65$. According to those criteria,  
 positions SW\#1 through \#7 classify as DLAs candidates. This
lends support to the idea that DLAs occur (at least) in regions internal to galaxies and,
furthermore, that some of them are associated with disks both at high
and   low redshift, as predicted by state-of-the-art
simulations~\citep{Rhodin2019}. Moreover, the present arc 
positions classified as DLAs have also the widest velocity dispersions (most
of them are within our `disk' kinematical classification),
suggesting we are hitting a prototype DLA host~\citep[e.g.,][]{Ledoux2006,Neeleman2013}.

Finding DLAs out to $15$\,kpc ($> 0.1~R_{\rm vir}$) may be somewhat
surprising. Halo models predict columns in excess of the DLA threshold only at
very low impact parameters, about three times less than
here~(\citealt{Qu2018}; but see \citealt{Mackenzie2019}). The larger extent
observed here might be due to the geometrical effect of probing along the
major axis of an inclined disk~\citep[but see][]{Rao2013}.

Assuming G1 hosts DLA clouds with unity covering factor
within a projected disk of radius 15\,kpc, we
estimate the total mass in neutral gas to be roughly  $\log M_{\rm HI}/M_{\astrosun} \approx
9.5$. This is of the order of magnitude of what is found
in 21-cm observations at low redshift~\citep[e.g.,][]{Kanekar2018},
suggesting that G1 represents a high-redshift analog of a nearby DLA host.

G1's star-formation efficiency, defined as SFR/$M_{\rm HI}$, is relatively
high, SFE=$3.5\times10^{-10}$\,yr$^{-1}$, for the bulk of star-forming
galaxies~\citep[][]{Popping2015}. On the other hand, the cool gas fraction,
defined as $M_{\rm HI}/(M_{\rm HI}+M_*)$, falls just below average for
$z=0.7$: $f_{\rm gas}\approx0.4$~\citep[e.g.,][]{Popping2015}. This indicates
that G1 is still efficiently forming stars, but will enter a quenching phase
---running out of gas in (SFE)$^{-1}\approx 3$\,Gyr--- if not provided with
extra gas supply~\citep{Genzel2010,Leroy2013,Sanchez2014}.

\subsection{Cold accretion}
\label{sec_accretion}

\begin{figure}
\includegraphics[width=\columnwidth,angle=0,clip]{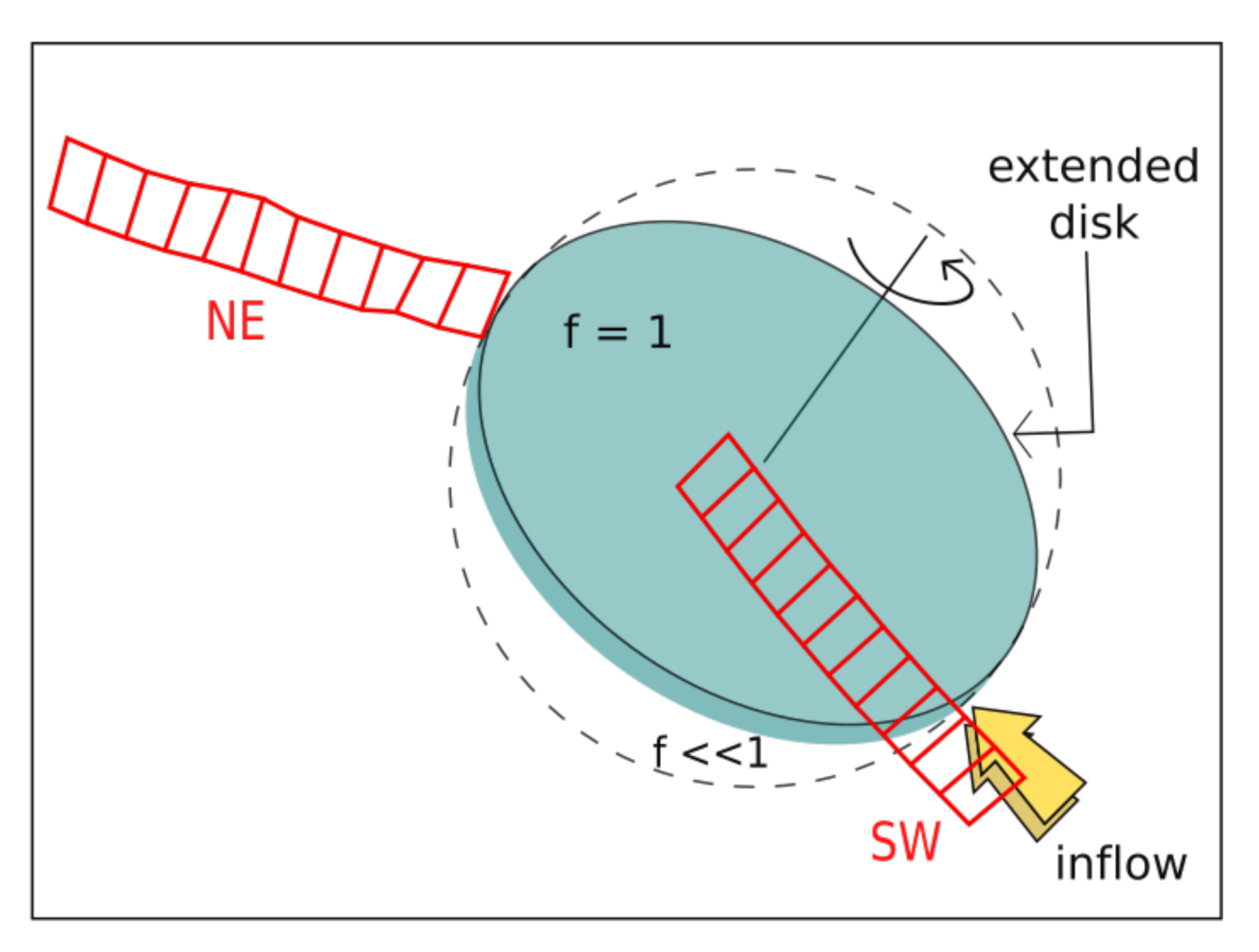}
\caption{Cartoon model for the inner CGM of the $z=0.7$ galaxy studied in this
  work (G1). The red polygons represent the MagE spaxels, reconstructed in the
  absorber plane and shown here in the same scale as in
  Fig.~\ref{fig_rEW_map}. The green rotating disk represents the volume where
  we detect \mgii\  absorption with $W_0>0.12$ \AA.  The disk is centered on the stellar light of G1, has
  a position angle of $55\degree$ N to E, and has an inclination angle of
  $i=45\degree$, i.e., same parameters as for the stellar disk (see also
  Fig.~\ref{fig_rEW_map}).  The disk is assumed
  to produce absorption with unity covering factor and to be embedded in a
  spherical volume producing much less covering at our detection limit. 
  The extensions of disk and spherical envelope are set
  arbitrarily such that no absorption is detected on spaxel NE \#11
  (right-most position in the NE slit). 
  The yellow arrow symbolizes in-flowing enriched gas
  which, if co-planar and aligned with the major axis, would reproduce the observed 
  \mgii\ kinematics at SW \#10 and SW \#11 (left panel in Fig.~\ref{fig_vel_D}). See \S~\ref{sec_accretion} for further discussion. 
\label{fig_cartoon}}  
\end{figure}

The \mgii\ gas detected at SW positions \#10 and \#11 stand out in many
respects (Fig.~\ref{fig_ratios}): it is kinematically detached from the
rotation curve; it has larger $W_0$ than an extrapolated trend followed by the
more internal positions; and it has the lowest \rfemg\ values, likely
indicating less processed gas. In addition, spaxel SW \#11 lies $7$\,kpc away
in projection from the major axis; depending on the (unknown) disk thickness,
the gas detected in these directions could be co-planar and lie at distances
of $\approx 0.2\,R_{\rm vir}$ from G1. These signatures suggest an 'external'
origin.  The absorption profiles at some other SW positions allow for an
unresolved velocity component at the velocity of SW \#11
(Figures~\ref{fig_stack_mage} and~\ref{fig_vel_D}), which could be explained
by extended non-rotating gas surrounding the disk.  However, such a velocity
component would not fit SW \#5 through SW \#9, nor any of the NE spaxels.  We
therefore dismiss the surrounding gas scenario for SW \#10 and \#11.  Rather,
we consider in-falling gas.  Cosmological simulations predict that galaxies
hosted by M$\lesssim 10^{12}$\,\msun~halos should undergo ``cold-mode''
accretion~\citep[e.g.,][]{Stewart2011a}.  In the following we consider the
possibility to have detected enriched cold accretion at medium
redshift~\citep{Kacprzak2014,Stewart2011,Bouche2013,Bouche2016,Danovich2015,Qu2019}.

Fig.~\ref{fig_cartoon} shows a cartoon representation of G1's inner CGM. The
green rotating disk represents the volume where we detect
\mgii-\feii-\mgi\ absorption.  The disk is assumed to produce absorption with
unity covering factor and to be embedded in a spherical volume likely
producing much less covering at our detection limit, $W_0>0.12$ \AA.  This
distinction is a possible explanation for the good match with an isothermal
model at the SW slit (Fig.~\ref{fig_rEW_D}) and the lack of detections at the
NE slit (Fig.~\ref{fig_rEW_map}).  In the cartoon model, the extensions of
disk and spherical envelope are set arbitrarily such that no absorption is
detected to the North-East of spaxel NE \#11 (right-most position in the NE
slit).  Such a choice implies that SW \#10 and SW \#11 (right-most positions
on the SW slit) would not have signal from the disk, but from an external
medium, which is consistent with our low-velocity detections. The proposed
accreting gas enters the galactic disk radially and roughly transversely to
the line-of-sight (producing the low line-of-sight velocities) while in the
process of acquiring enough angular momentum to start co-rotating.

Alone from the kinematics, though, it is hard to disentangle extraplanar
inflow (radial or tangential) from a warped disk~\citep{Diamond-Stanic2016}, a
scenario that seems to reproduce some observations of quasar absorbers having
low line-of-sight velocities~\citep{Rahmani2018,Martin2019}.  Indeed, most of
the \hi\ disks in the local Universe exhibit
warps~\citep{Sancisi2008,Putman2009}, their extended \hi\ disks do show
anomalies~\citep{Koribalski2018}, and in a few cases rotation curves start
declining when \hi\ becomes patchy in the extended disk of dwarf
galaxies~\citep{Das2019,Oikawa2014}.  Authors explain such cases via warped
and tilted disks~\citep{Sofue2016}.

This being said, our data offer enough indications {\it against} the warped
disk scenario. First, we do not see interacting galaxies
\citep{Diamond-Stanic2016}.  Secondly, we do not detect absorption at the same
distance on the opposite side of G1 (i.e., NE positions \#10 and \#11).  Third,
velocities in simulated dwarfs fall down by only 20\% at 20
kpc~\citep{Kyle2019}, while here we see a decline of about $80\%$.  Indeed, SW
positions \#10 and \#11 have much less specific angular momentum than the
rest. For instance SW \#11 has 60\% less specific angular momentum than SW
\#10 (i.e., $(Rv)_{\# 11 }=0.6\times (Rv)_{\#10}$), and so forth, suggesting
the gas is not (yet)  rotating. And lastly, the gas shows the lowest
\rfemg\ values, i.e., it is consistent with less processed gas, which is
expected in cold accretion~\citep[e.g.,][]{Oppenheimer2012,Kacprzak2016}. Detecting accretion via \mgii\ at the level
of $W_0\sim0.2$--$0.3$ \AA, although incompatible with pristine
gas~\citep{Fumagalli2011,MartinDC2019}, agrees well with quasar observations
of disk-selected absorbers~\citep{Rubin2012,Zabl2019}.

By averaging spatially the absorption in SW\#10 and \#11  in a circular aperture of
radius 30\,kpc, we find that the covering factor is low, $f_{\rm accretion} \approx
1$\%. This is consistent with  
simulations at higher redshifts~\citep{Faucher2011,Fumagalli2011} and lends
support to the cold-accretion scenario.

Altogether, cold, recycled accretion~\citep{Rubin2012,Danovich2015} at
$\approx0.2\,R_{\rm vir}$ seems the most favoured scenario to explain the
present data.  It might be radial accretion at the disk
edge~\citep{Stewart2011,Putman2012} originating from the cool
CGM~\citep{Werk2014} in form of recycled
winds~\citep{Oppenheimer2010,Angles-Alcazar2017}, i.e., gas left over from past
star-bursts.

This is not the first time absorption kinematics is seen decoupled from
emission~\citep{Steidel2002,Martin2019,Ho2017}.  Velocities below Keplerian
have also been detected in quasar sightlines although at slightly larger
distances~\citep{Martin2019, Ho2017,Kacprzak2017}. Those signatures seem to be
frequent in highly inclined disks and authors have argued that they might
probe inflows. However, with quasar sightlines probing only one position in
the intersected halo, it is challenging to confirm this hypothesis. Thanks to
the present tomographic data, we see for the first time a {\it smooth
  transition} to disk co-rotation, providing the first unambiguous evidence
for enriched-gas accretion beyond the local Universe.

\section{Summary and conclusions}\label{sec:summary}

We have studied the cool and enriched CGM of a $z=0.7$ star-forming galaxy
(G1) via the gravitational arc-tomography technique \citep{Lopez2018}, i.e.,
using a bright giant gravitational arc as background source. G1 appears to be
an isolated and sub-luminous disky galaxy, seen at an inclination angle
$i\approx 45\degree$.

We have measured \mgii, \feii, and \mgi\ equivalent widths ($W_0$) in 25
$3\times6$ kpc$^2$ independent positions (including 13 velocity measurements) along G1's
major axis, at impact parameters $D=0$--$60$\,kpc (0--0.4\,$R_{\rm vir}$). This
unique configuration has allowed us to probe distinct signatures of the CGM in an
individual galactic environment. Our findings can be summarized as follows:

\begin{enumerate}

\item Enriched gas is detected out to $D\approx 30$\,kpc ($\approx 0.2~R_{\rm vir}$) in one
radial direction from G1. The absorption profiles (Fig.~\ref{fig_stack_mage}) show kinematic
variations as a function of $D$, becoming less complex outwards of G1. We suggest that
the arc positions probe different regions in the halo and extended disk of G1. Within
$\sim 3$~kpc, the smallest scales permitted by our ground-based observations, the
gas distribution appears smooth in the central regions (unity covering factor). By
comparing $W_0$ measured on both sides of G1, we find evidence that the gas is not
distributed isotropically (Fig.~\ref{fig_rEW_map}).

\item We observe a $W_0$--$D$ anti-correlation in all three studied metal species.
The $W_0(2796)$ scatter in the arc data (Fig.~\ref{fig_rEW_D}) is significantly smaller than
that of the quasar statistics, suggesting biases in the latter, likely due to a variety of
host properties and orientations. Our data populates the sparse $D<10$\,kpc
interval, revealing that $W_0(D)$ flattens at low impact parameters. An isothermal density profile
fits the arc data remarkably well at almost all impact parameters. Since most of the model
parameters are tied to the quasar statistics, this suggests that the present halo is
prototypical of the \mgii-selected CGM population. In particular, at $D<10$\,kpc the good
fit rules out cuspy gas distributions, like those described by NFW
or power-law models.

\item For most of the detections, the absorption velocities  
(Fig.~\ref{fig_vel_D}, left panel) resemble a flat rotation curve, which appears to be
kinematically coupled to G1's \oii\ emission. There are two exceptions to this trend. (a) 
One position, lying only $4$\,kpc in projection from G1 and measured independently in two
slits, departs from rotation with a velocity of $\sim +90$\,\kms. This suggests that the
gas, also exhibiting the highest \rfemg\ value of the sample, might be out-flowing from G1.
And (b), the two outer-most detections (at $\approx 30$~kpc $\approx0.2~R_{\rm vir}$) also seem
decoupled from the disk kinematics, falling too short in velocity. We do not detect
absorption at the same distance on the opposite side of G1. We interpret the low-velocity signal  as occurring in less-enriched gas having a co-planar trajectory, which will
eventually flow into the galaxy's rotating disk (e.g., an enriched cold-accretion inflow).

\item The equivalent-width ratio \rfemg$(D)$ (Fig.~\ref{fig_ratios}) exhibits a negative
gradient, which could partly be due to a negative gradient in metallicity. This ratio
also suggests that G1's central regions ($D<15$\,kpc) may host DLAs. We estimate the
total reservoir of neutral gas and find it to be comparable with the mass locked into
stars, suggesting that the galaxy has little fuel left to keep up with its
current star-formation efficiency.

\end{enumerate}

\section{Outlook}\label{sec:outlook}

We have highlighted the exquisite advantages of gravitational arc-tomography: (1) the
background sources extend over hundreds of kpc$^2$ on the sky, permitting a true
`slicing' of the CGM of {\it individual} intervening galaxies; 
(2) comparison with the statistics of quasar-galaxy pairs offers a great opportunity to
assess the gas patchiness and its covering factor around individual systems, something
beyond the capabilities of present-day quasar observations; (3) the individual systems
can be used as test laboratories in future simulations. 
Challenges are manyfold as well: sensitive spatially-resolved spectroscopy is needed (not
available until recently); absorber-plane reconstruction is required via ad-hoc modeling
of the lensing configuration (usually non-trivial); bright giant gravitational arcs are
rare on the sky. We expect that soon new surveys will provide targets for
future extremely-large observing facilities. In the meantime, a comparison scheme between
the arc and quasar statistics can and must be developed.
These are key aspects that nicely {\it complement} quasar studies. Furthermore, with higher
spectral resolution one shall be able to resolve individual velocity components and
assess the chemical state of the gas in a spatial/kinematical context. Undoubtedly, such
tools shall enable a more profound understanding of the baryon cycle across
galaxy evolution.

\section*{Acknowledgements}

We thank the anonymous referee for comments that improved the manuscript.
This work has benefited from discussions with Nikki Nielsen, Kate Rubin, Glenn
Kacprzak, Umberto Rescigno and Nicolas Bouch\'e. This paper includes data
gathered with the $6.5$ meter Magellan Telescopes located at Las Campanas
Observatory, Chile: the Magellan/MagE observations were carried out as part of
program CN2017B-57 (PI Tejos). The VLT/MUSE data were obtained from the ESO
public archive (program 297.A-5012(A), PI Aghanim).  This work was supported
in part by NASA through a grant (HST-GO-15377.01, PI Bayliss) awarded by the
Space Telescope Science Institute, which is operated by the Association of
Universities for Research in Astronomy, Inc. under NASA contract NAS 5–26555.
SL was partially funded by UCh/VID project ENL18/18 and by FONDECYT grant
number 1191232. NT acknowledges support from PUCV/VRIEA projects
$039.333/2018$ and $039.395/2019$, and FONDECYT grant number 1191232.  LFB was
partially supported by CONICYT Project BASAL AFB-170002. MG was supported by
NASA through the NASA Hubble Fellowship grant \#HST-HF2-51409 awarded by the
Space Telescope Science Institute, which is operated by the Association of
Universities for Research in Astronomy, Inc., for NASA, under contract
NAS5-26555.

\vspace{5mm}

\bibliographystyle{mnras}
\bibliography{Lopez_lit}

\appendix

\section{Magellan/MagE data}
\label{sec:mage_appendix}

\subsection{Magellan/MagE data acquisition}
\label{sec:mage_acq}

We observed the two northernmost segments in \pks\ with
Magellan/MagE \citep{Marshall2008} in dark-time on July 20th and 21st, 2017
(both first half-nights). Given the large declination of the field the
observations were conducted with airmass restricted between $1.55$ and
$1.66$. The seeing was good and steady on both half-nights, varying between
$0\farcs6$ and $0\farcs7$. However, the general weather conditions varied
between the two half-nights, being cloudy (cirrus) during most of the first
half-night and clear on the second. This situation affected the quality of
some of the exposures taken during the first half-night, from which only one
exposure of the SW slit was finally used. On the other hand, we used all
exposures taken for the NE, SW and SKY slits on the second
half-night. Table~\ref{tab:obs} gives a summary of the useful observations.

Data acquisition was performed from blind offsets with respect to a nearby
bright star located at Celestial Coordinate (J2000) R.A.$=$ 15h\,50m\,00s and
Dec. $=-78$\degree\,10m\,57s.  Because our PAs are different than the
Parallactic Angle at a given time (\PAne, \PAsw and \PAsky, for `NE', `SW'
and `SKY', respectively), we used a blue-filter in the acquisition camera; in
this manner, we ensure that the bluest possible optical coverage of the arcs
(where the transitions of interest fall), were correctly aligned within the
slits.

Individual exposure times varied between $2\,700$--$4\,500$\,s, for a total of \netot, \swtot\ and \skytot\ hours for slits `NE', `SW' and `SKY', respectively, as presented in Table~\ref{tab:obs}.

\subsection{Magellan/MagE data reduction}
\label{sec:mage_reduction}

We used a custom pipeline to reduce the MagE data. Because we deal here with an extended source, the main task is to account for the known misalignment
between the spatial direction and the CCD columns on the two-dimensional (2D) spectra, 
a tilt which varies with position on the 
CCD~\citep[e.g.,][]{Bochanski2009}. To this end, for each 2D spectrum we
define 33 one-pixel ($0\farcs3$) long 
pseudo-slits, whose positions correspond to an offset with respect to the previously defined echelle orders, and for each pseudo-slit we obtain independent wavelength solutions. Variance frames are created from the 2D spectra of each object, and cosmic rays are assigned with `infinite' variances.

We extract and reduce the flux and the variance at each pseudo-slit by
 linearly interpolating the values in the image, order by order.  A master sky
 spectrum is obtained from the slit `SKY' by averaging spatially 20
 pseudo-slits in the central part of the slit (the slit extremes have flux
 from the arc segments by design).  A scaled version of the master sky is
 subtracted equally to the flux at all pseudo-slits in a given exposure, with
 scale factors chosen in such a way that SW positions \#2, \#3 and \#4 (all
 having black absorption) end up with no residuals above the zero flux level.

A response function is created by reducing a spectro-photometric
standard. Division of the object spectra by this function converts the
sky-subtracted counts into flux units and corrects the blaze function of the
spectrograph. The orders are merged into a one-dimensional, calibrated and
reduced spectrum. Finally, wavelengths are corrected for barycentric
velocities and the different exposures co-added optimally by weighting by the
inverse variances.

As a compromise between matching the seeing and maximizing S/N, for each of the `NE', `SKY', and `SW' slits, we combine the 33  pseudo-slit spectra along $3$ consecutive offsets (Fig.~\ref{fig_2D}).  This spatial binning defines $11$ `pseudo-spaxels' on the plane of the sky (referred to as `MagE spaxels' or `positions') of $0\farcs9\times1\farcs0$ each, oriented along the slits (Fig.~\ref{fig_FOV}, Fig.~\ref{fig_slit_sky} and subsequent figures). Each spaxel is separated by $\gtrsim 1$ seeing units from
the next, ensuring that the signals are mostly independent.  Due to the inhomogeneous source brightness along the slit (and not due to partial source illumination; see Fig.~\ref{fig_slit_sky}), the spectra have different S/N, ranging typically from 4 to 10.

The final resolving power, as measured from sky emission lines, is $R=4\,500$
with a dispersion of $0.37183$\,\AA\,pix$^{-1}$ (or $\approx 22$\,\kms\ at the
position of \mgii) and a RMS of $\approx 0.06$\,\AA. This RMS is similar to
that reported by~\cite{Bochanski2009}.  To check the wavelength calibration,
we select sky lines in the MagE and in the MUSE data
(see \S~\ref{sec:muse_data}) and calculate their centroids. An histogram of
velocity differences appears centered around zero with a dispersion of
$\sigma=12$\,\kms; therefore, these reduced MagE spectra can be compared with
the MUSE spectra. As a sanity check, we inspected the match with the sky line
at $\lambda=4861.32$ \AA\ (this is right at the position of the
expected \mgii\ absorption; Fig.~\ref{fig_2D}) and found them to be
consistent.

Finally, the combined spectra for a given slit were recorded into data-cubes
of a rectangular shape of $1 \times 11$ spaxels. Throughout the paper we use
the convention that the northernmost spaxel in a given slit is `position 1'
(\#1) and these increase towards the South in a consecutive order, being the
`position 11' (\#11) the southernmost spaxel in a given MagE slit
(e.g. "SW \#1, \#2, \#3, etc.; see Figs.~\ref{fig_FOV} to~\ref{fig_2D}).

\subsection{Magellan/MagE astrometry}
\label{sect_astrometry}

Although the slit acquisition was executed by a blind offset from a reference
source, the process of acquiring the reference star was performed manually (by
the telescope operator) and may introduce a small position offset of a
fraction of an arc-second. To make sure that the astrometry of our MagE
slit data-cubes matches to that of MUSE (and hence {\it HST}; see
Section~\ref{sec:muse_data}), we proceed as follows.

First, we used the MUSE data-cubes as reference to create several mock MagE
data-cubes from the MUSE data (referred to as `MUSE-MagE') using the PyMUSE
package \citep{Pessa2018}. Each MUSE-MagE data-cube has $11$ spaxels with
the exact geometry of our MagE slits, placed at a fixed PA (given by the
corresponding MagE slit) but with a different central position. As we made
sure the arcs segments were well within the slits, we only varied the slits
positions over $\pm 1$\arcsec {\it along} their corresponding PA
directions. 

For each of the resulting MUSE-MagE data-cubes, we compared the
spectral shape and total flux per spaxel to those of the actual corresponding MagE data-cube
(rebinned to the coarser wavelength dispersion of MUSE) within the wavelength
range between $6420-6440$\,\AA, i.e. encompassing the
\ciii~$\lambda\lambda1907,1909$ emission line of the arc source.\footnote{The
  \ciii\ emission line of the source is useful in this context because it
  appears as several unresolved knots at different relative positions along
  each arc segment.} We computed a spectral and a total flux (per spaxel)
  $\chi^2$ for the different mock MUSE-MagE datacubes and adopted the position
  that minimized it as our astrometry solution (as presented in
  Fig.~\ref{fig_FOV}). From the shape of the $\chi^2$ curves around the
  minimum we estimate a position uncertainty of the MagE slits of
  $0.2$\arcsec. We note that this is a systematic uncertainty that applies to
  {\it all} MagE spaxels in the same direction for a given MagE slit.

\section{Morpho-kinematical analysis of G1 from [O~II] emission}
\label{sec:galpak}

We performed a morpho-kinematical analysis of G1 using the \galpak\
software \citep[v.1.11;][]{Bouche2015}, from which we fit a rotating disk
model to the \oii\ emission observed in the MUSE datacube. First, we created a
de-lensed MUSE datacube at the absorber plane, centered around G1. We
re-sampled the resulting smaller de-lensed spaxels into the MUSE pixel scale
of $0.2$\arcsec per pixel, thus preserving the original geometry.  The input
instrument line-spread function was that of MUSE (FWHM=2.675\,\AA), and the
input PSF was the effective reconstructed (elongated) seeing PSF at the
absorber plane.

We run \galpak\ on the continuum-subtracted and de-lensed datacube around
the \oii\ emission (centered around G1), for $10\,000$ iterations until
convergence for a disk model with the following parameters: an exponential
flux profile, an arctan rotation curve, and a Gaussian thickness
profile. Although the \oii\ flux profile was not properly modeled as a single
disk component (some significant residuals are present due to the presence of
clumpy star-formation regions), the model did converge to a satisfactory
kinematical solution whose main parameters are presented in
Table~\ref{table_G1}.  We finally estimated a virial radius, $R_{\rm vir}$,
and a dynamical halo mass, $M_h^{\rm dyn}$, assuming a spherical collapse
model as $R_{\rm vir} = 0.1 H(z)^{-1} v_{\rm max}$, and $M_{\rm dyn}= 0.1
H(z)^{-1}G^{-1}v_{\rm max}^3$, where $H(z)$ is the Hubble parameter at
redshift $z$, and $G$ is the gravitational constant \citep[][see
Table~\ref{table_G1}]{Mo1998,Rahmani2018}.

\clearpage

\section{Fitted absorption-line profiles}

\begin{figure*}
\begin{minipage}{2.\columnwidth}
\includegraphics[width=0.22\columnwidth,trim={0cm 0cm -0.3cm 0cm}]{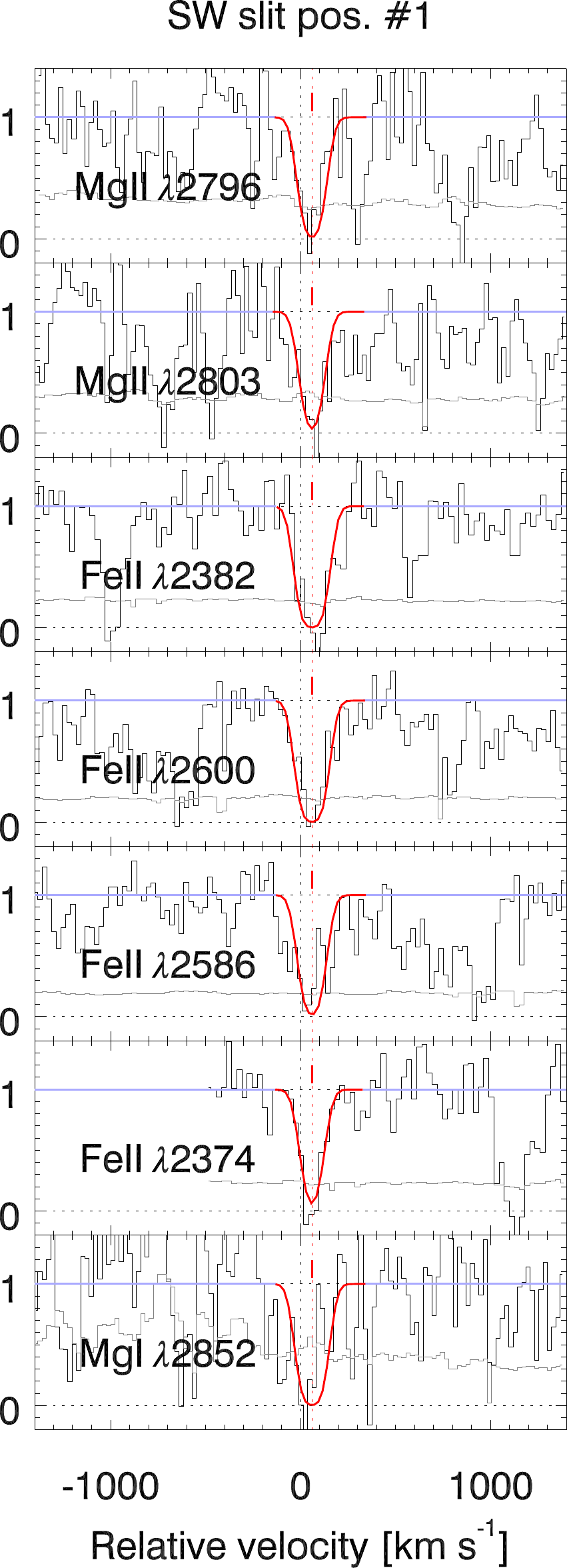}
\includegraphics[width=0.22\columnwidth,trim={0cm 0cm -0.3cm 0cm}]{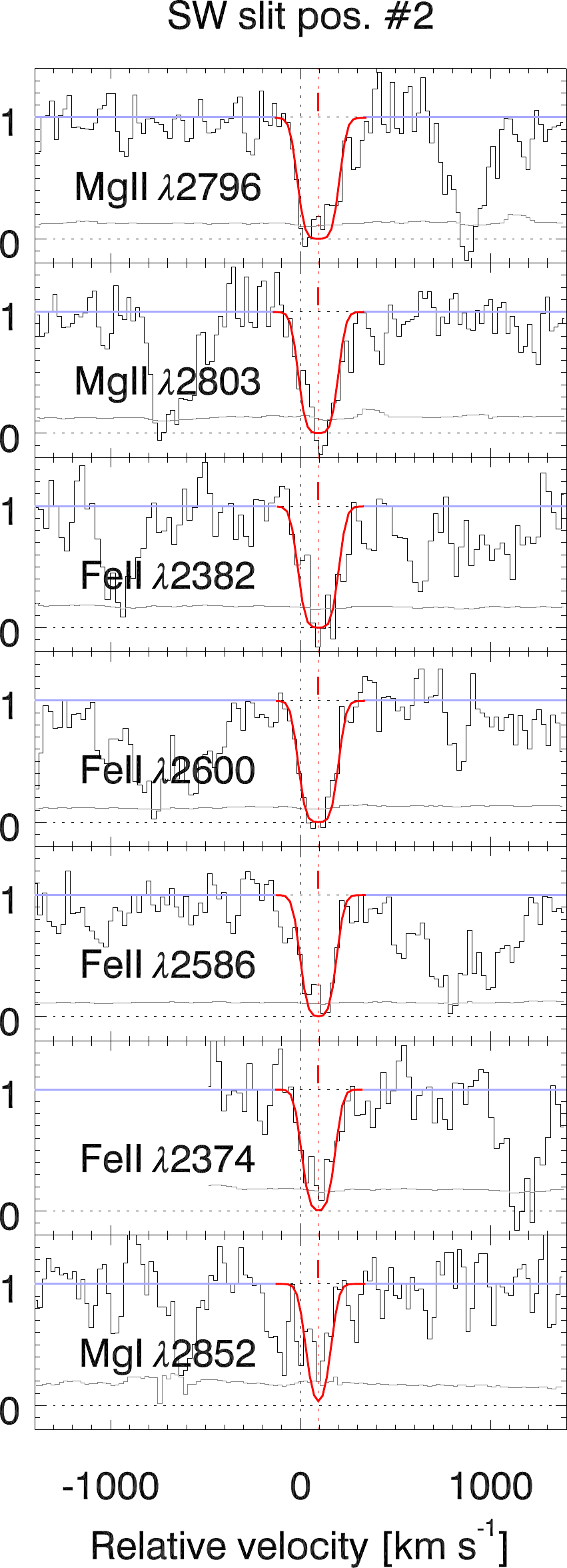}
\includegraphics[width=0.22\columnwidth,trim={0cm 0cm -0.3cm 0cm}]{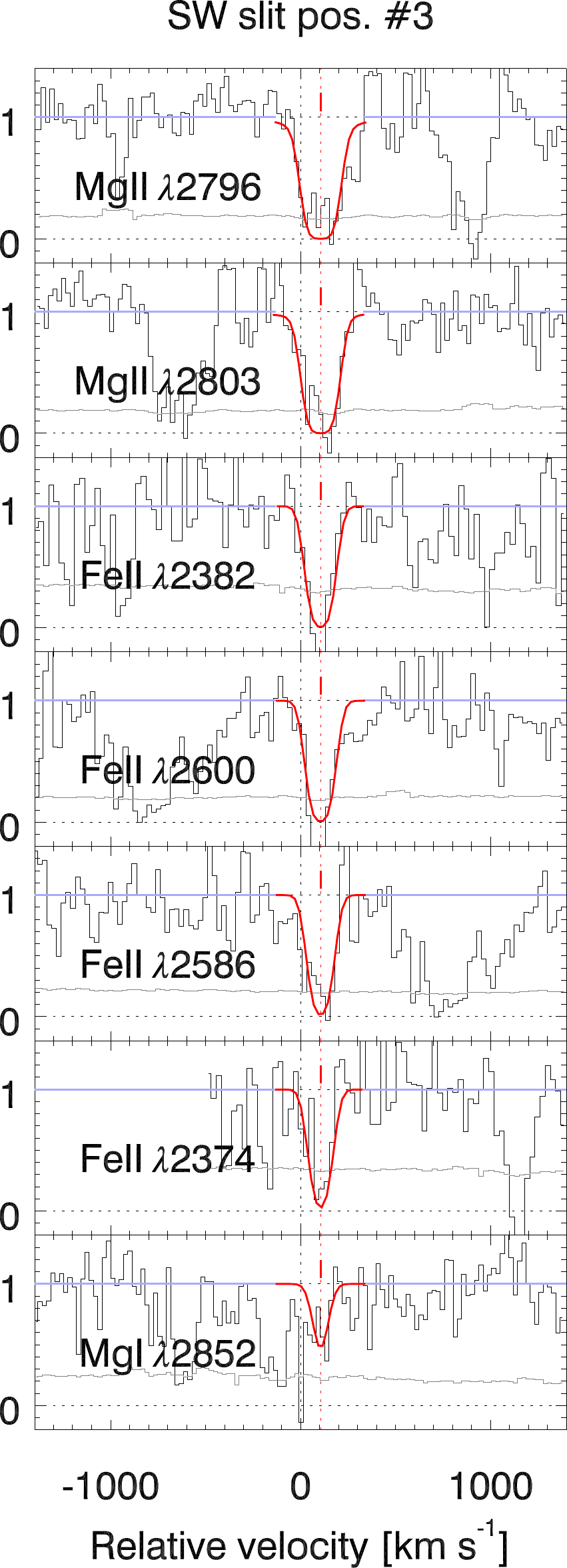}
\includegraphics[width=0.22\columnwidth,trim={0cm 0cm -0.3cm 0cm}]{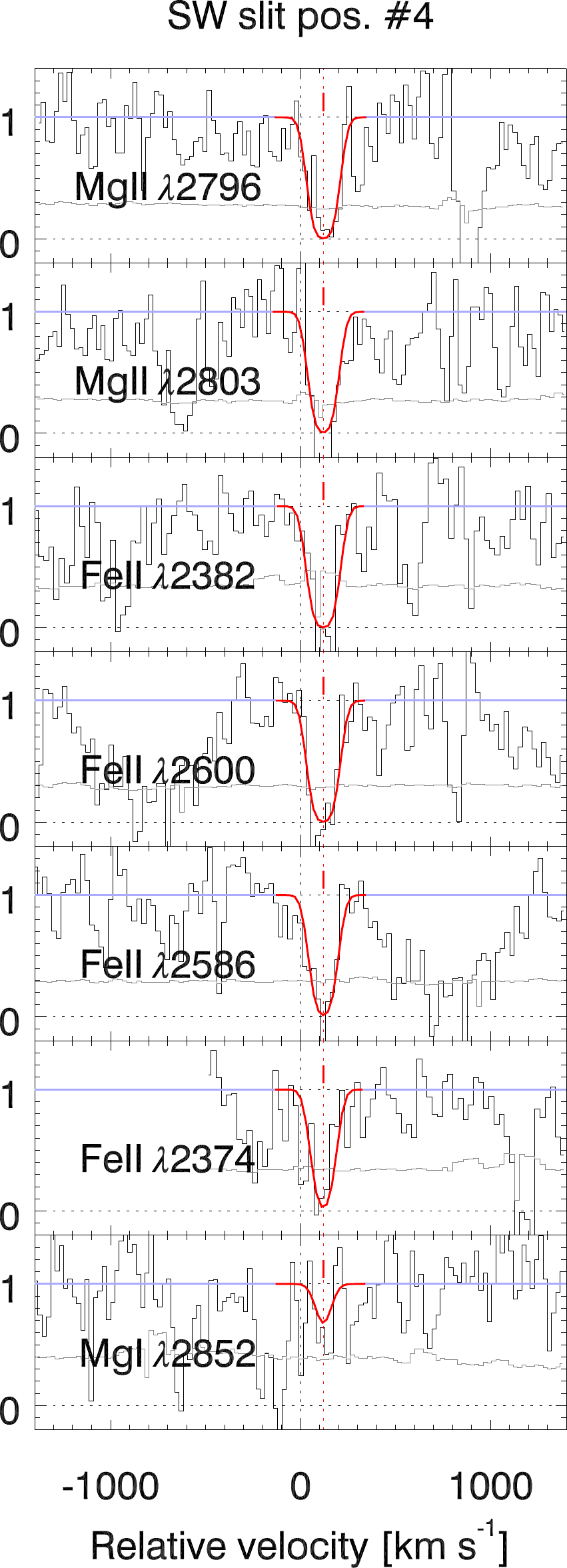}
\end{minipage}
\caption{Transitions detected in SW positions \#1 through \#4. The histograms show
the normalized flux and its $1\sigma$ error. The red curves are the fitted Voigt profiles.}
\label{voigt_profiles}
\end{figure*}

\begin{figure*}
\begin{minipage}{2.00\columnwidth}
\includegraphics[width=0.22\columnwidth,trim={0cm 0cm -0.3cm 0cm}]{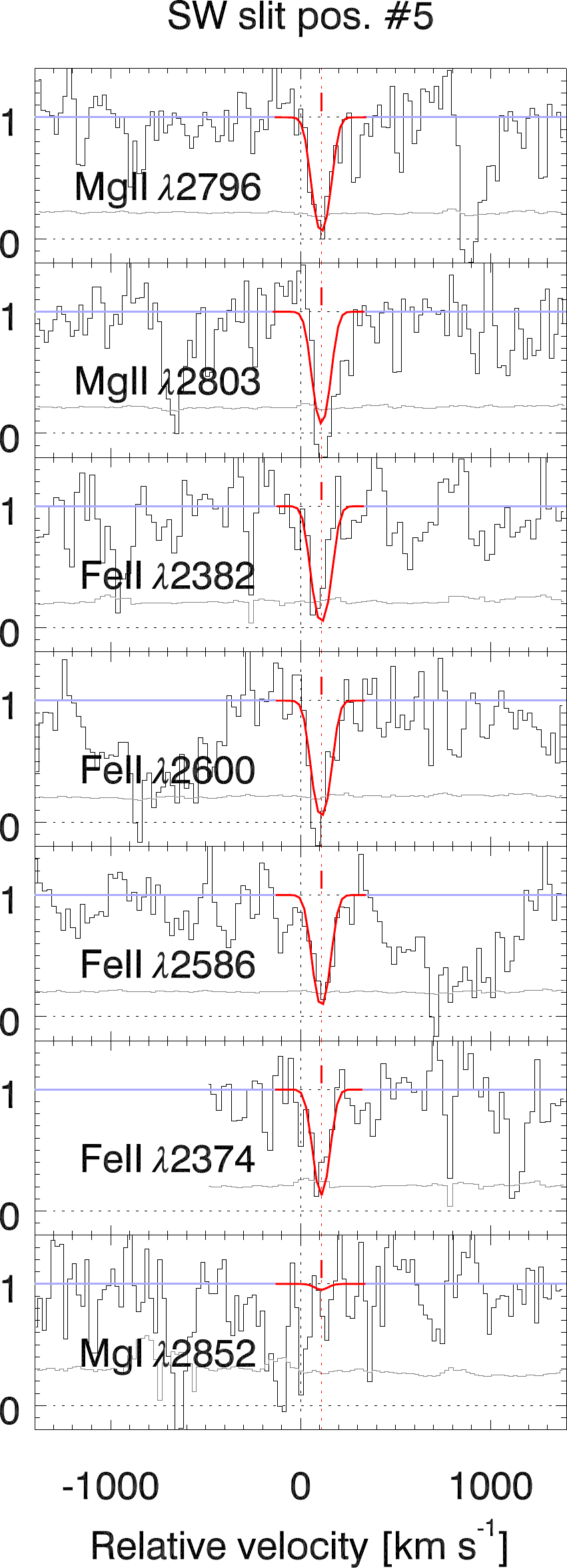}
\includegraphics[width=0.22\columnwidth,trim={0cm 0cm -0.3cm 0cm}]{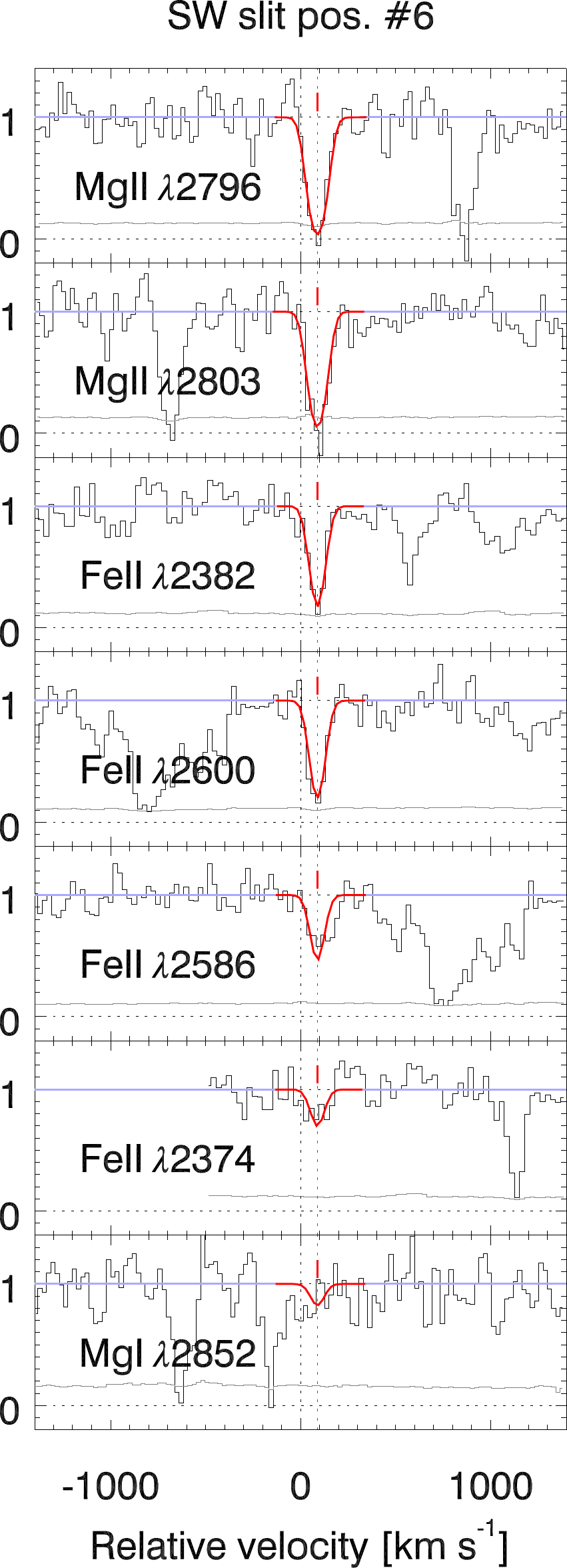}
\includegraphics[width=0.22\columnwidth,trim={0cm 0cm -0.3cm 0cm}]{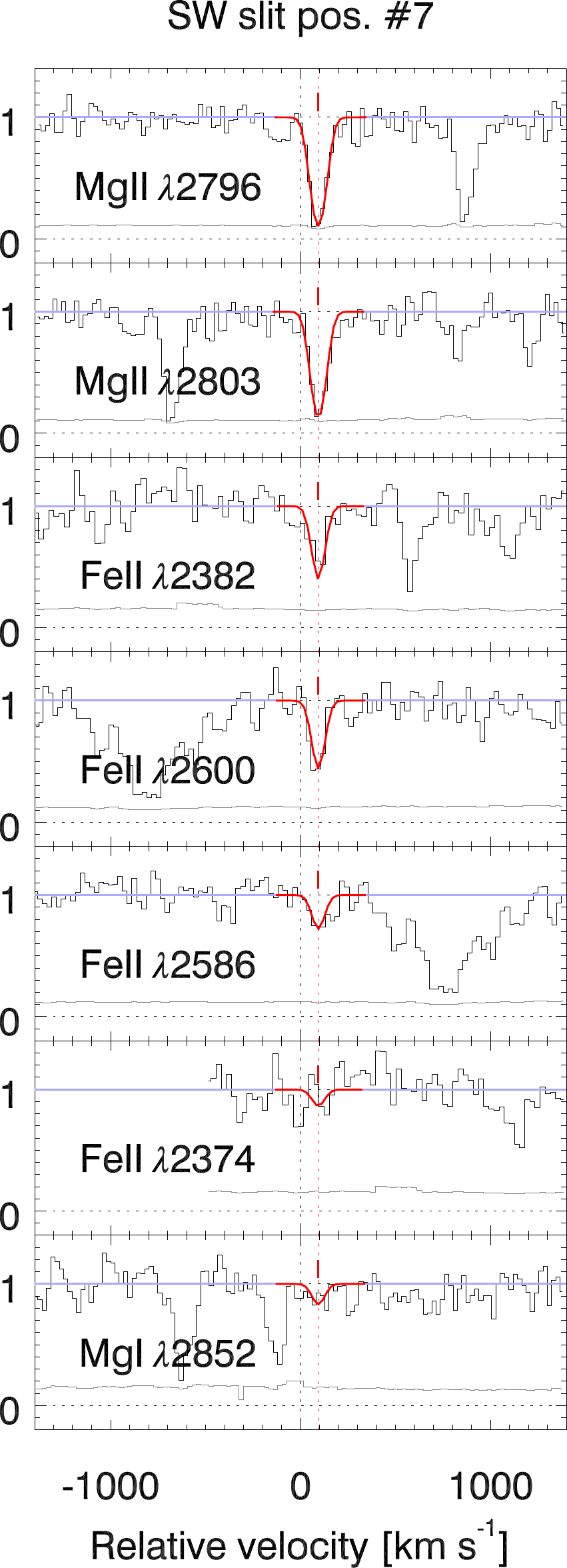}
\includegraphics[width=0.22\columnwidth,trim={0cm 0cm -0.3cm 0cm}]{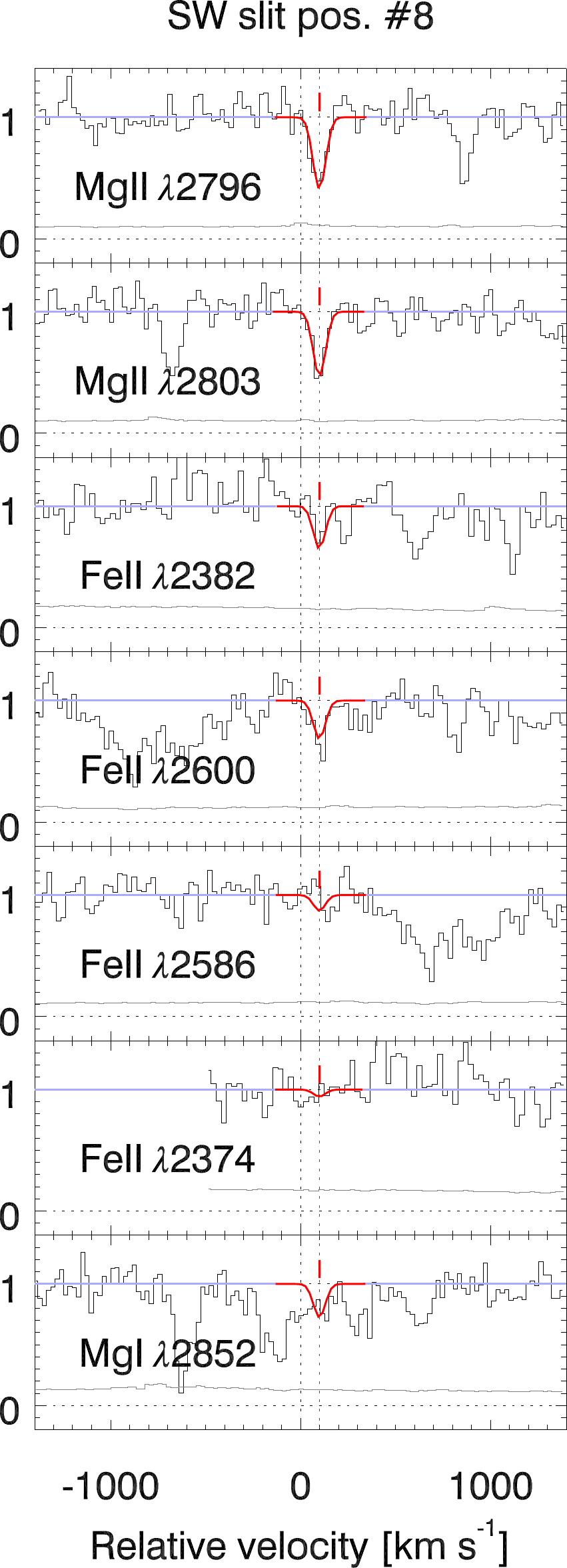}
\end{minipage}
\caption{Same as Fig.~\ref{voigt_profiles} for SW positions \#5 through \#8.}
\end{figure*}

\begin{figure*}
\begin{minipage}{2.00\columnwidth}
\includegraphics[width=0.22\columnwidth,trim={0cm 0cm -0.3cm 0cm}]{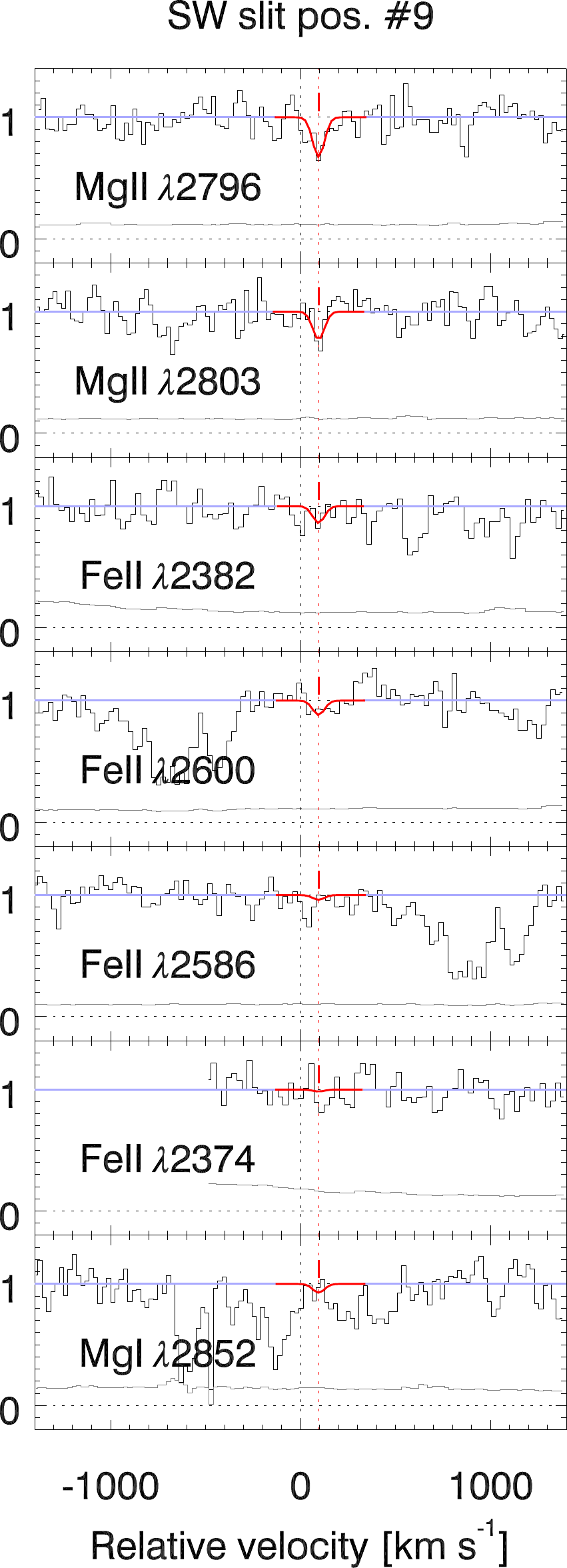}
\includegraphics[width=0.22\columnwidth,trim={0cm 0cm -0.3cm 0cm}]{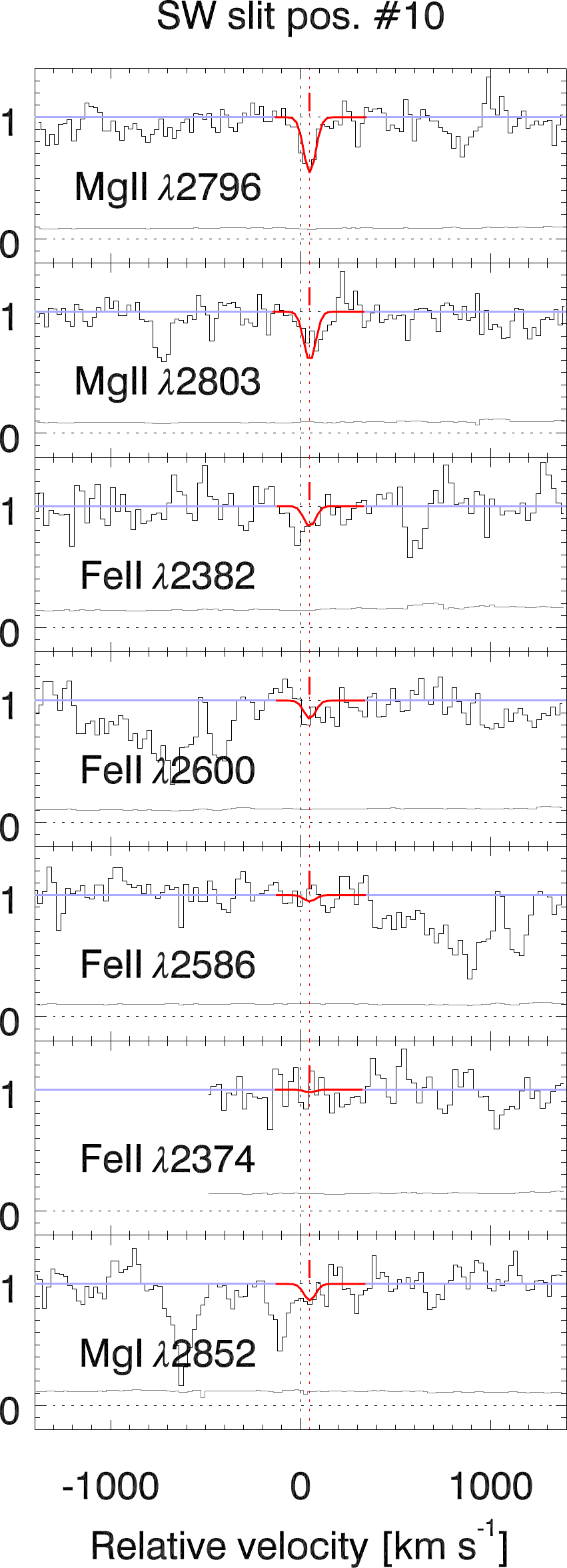}
\includegraphics[width=0.22\columnwidth,trim={0cm 0cm -0.3cm 0cm}]{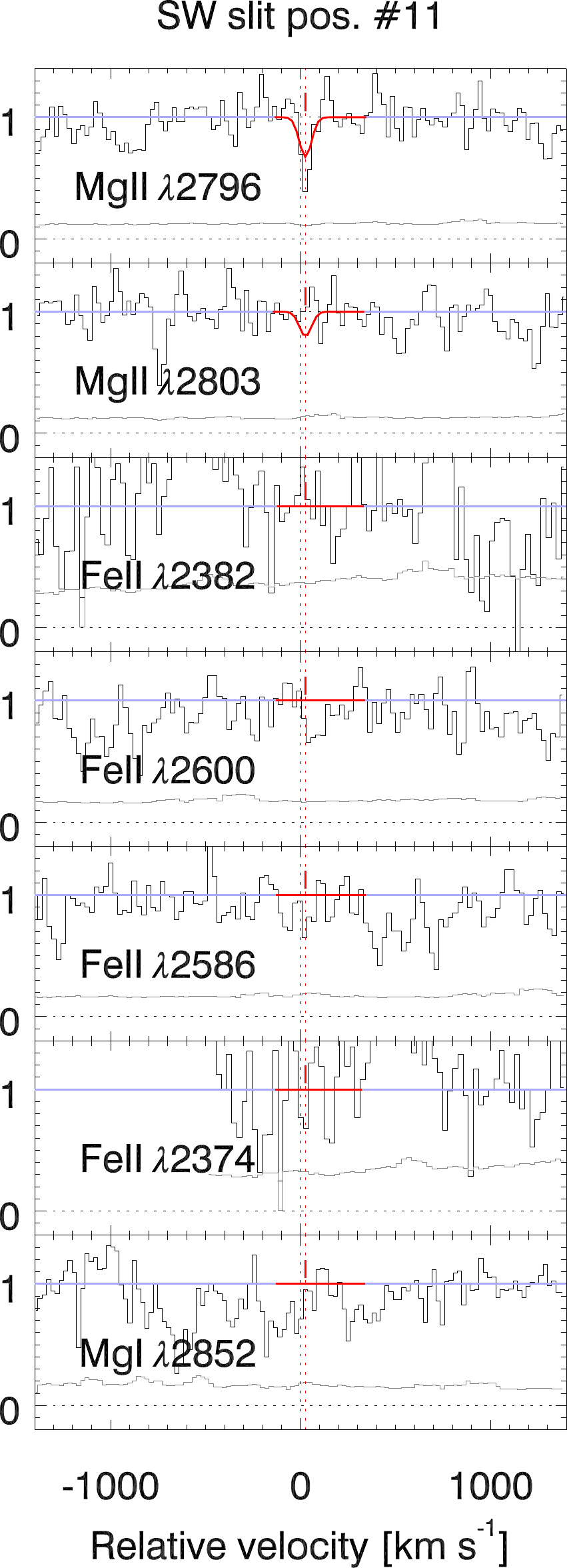}
\end{minipage}
\caption{Same as Fig.~\ref{voigt_profiles} for SW positions \#9, \#10, and \#11.}
\end{figure*}

\begin{figure*}
\begin{minipage}{2.00\columnwidth}
\includegraphics[width=0.22\columnwidth,trim={0cm 0cm -0.3cm 0cm}]{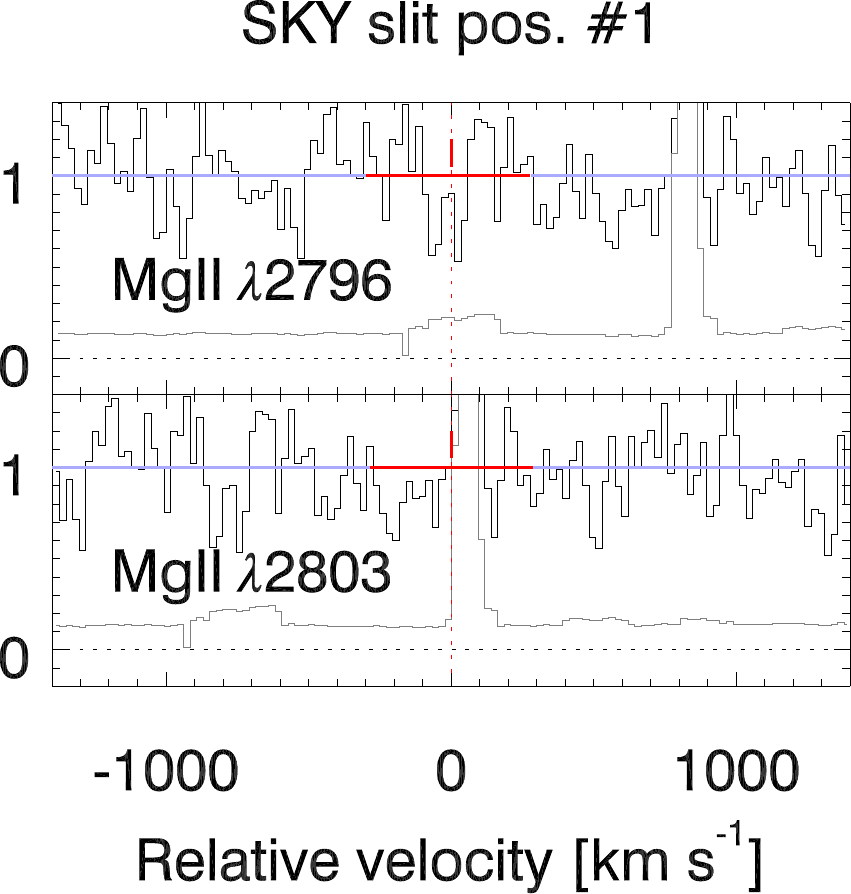}
\includegraphics[width=0.22\columnwidth,trim={0cm 0cm -0.3cm 0cm}]{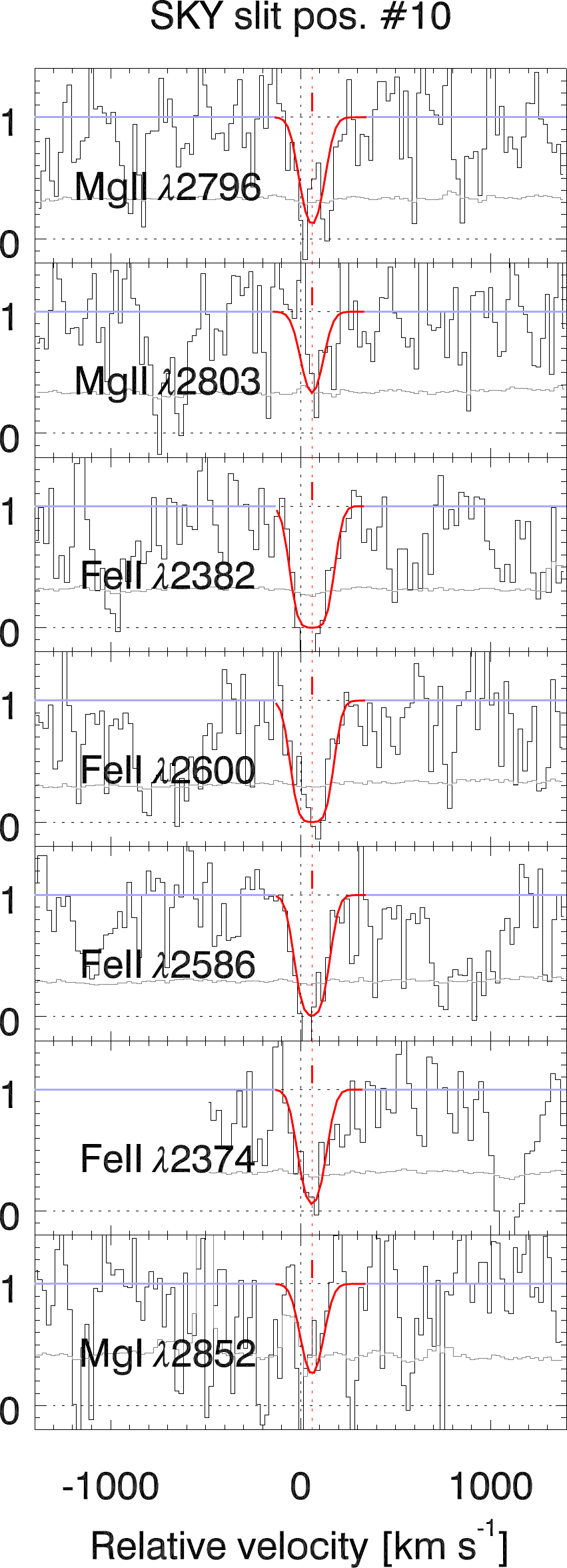}
\includegraphics[width=0.22\columnwidth,trim={0cm 0cm -0.3cm 0cm}]{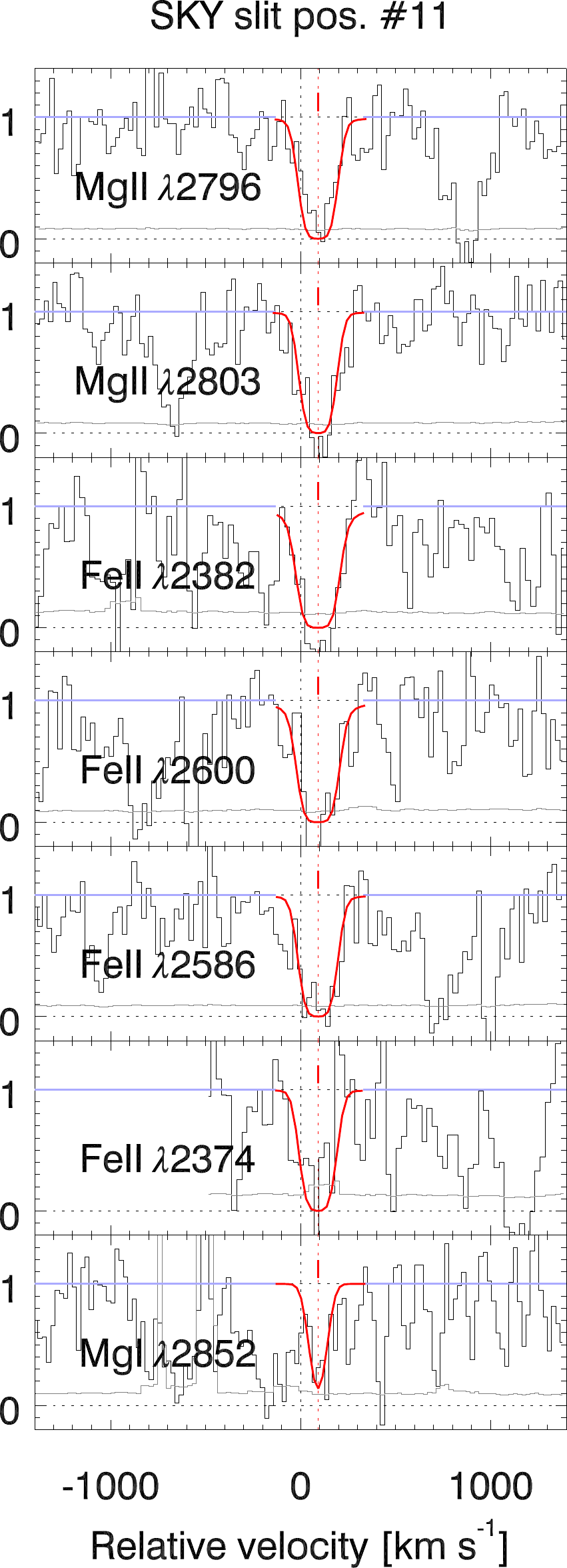}
\end{minipage}
\caption{Same as Fig.~\ref{voigt_profiles} for SKY positions \#1, \#10, and \#11.}
\end{figure*}

\begin{figure*}
\begin{minipage}{2.00\columnwidth}
\includegraphics[width=0.22\columnwidth,trim={0cm 0cm -0.3cm 0cm}]{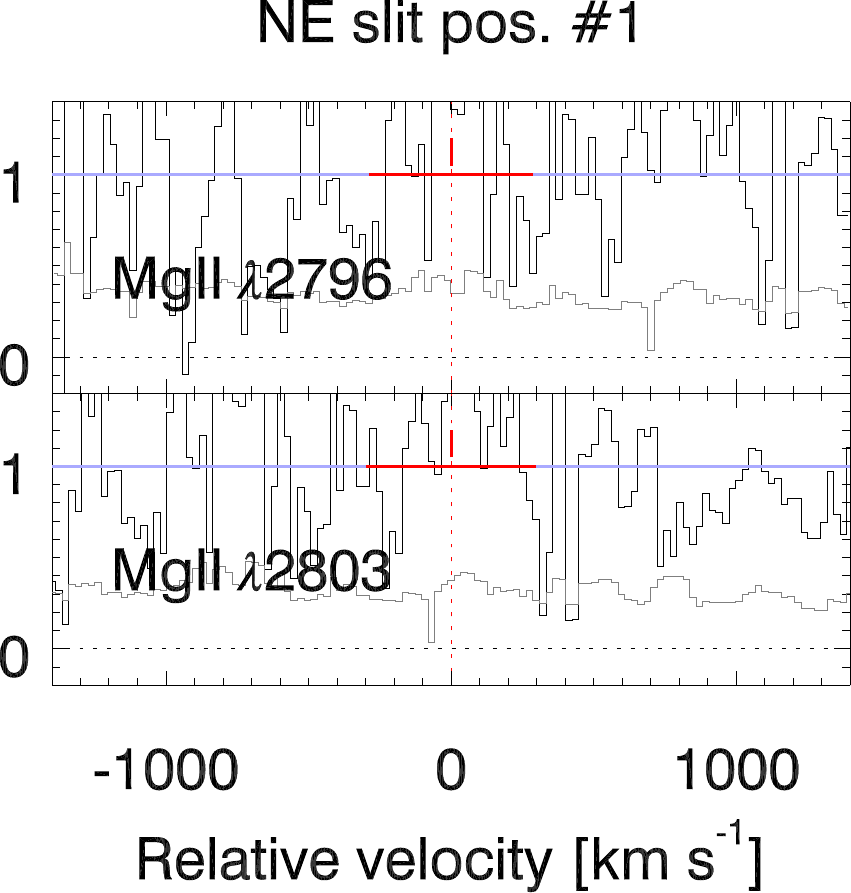}
\includegraphics[width=0.22\columnwidth,trim={0cm 0cm -0.3cm 0cm}]{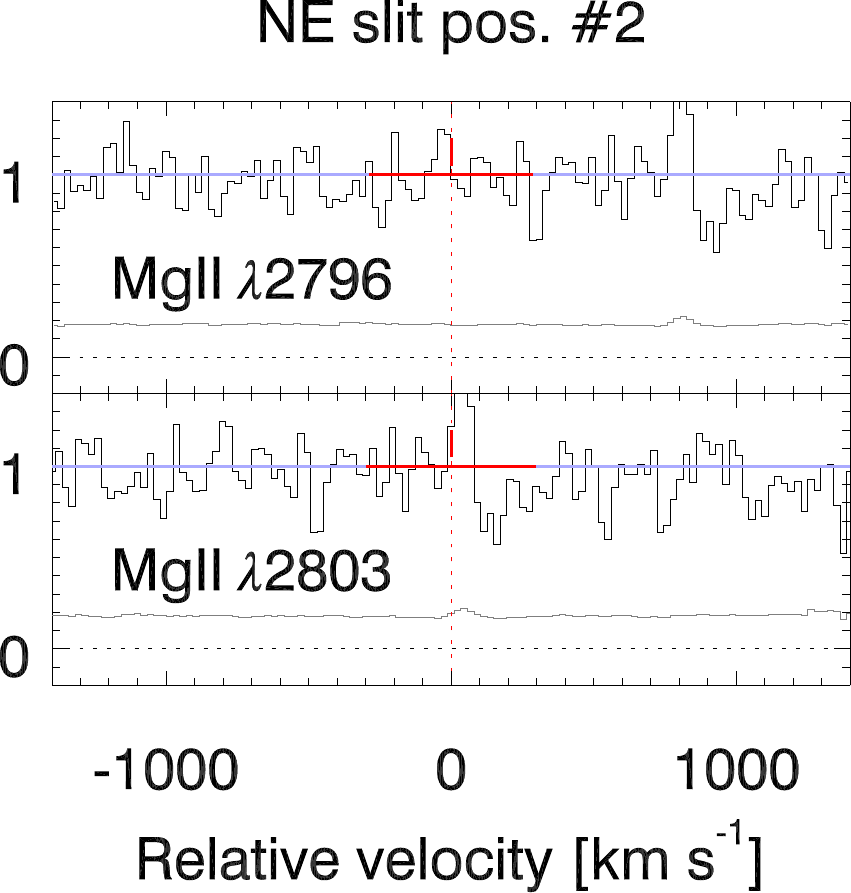}
\includegraphics[width=0.22\columnwidth,trim={0cm 0cm -0.3cm 0cm}]{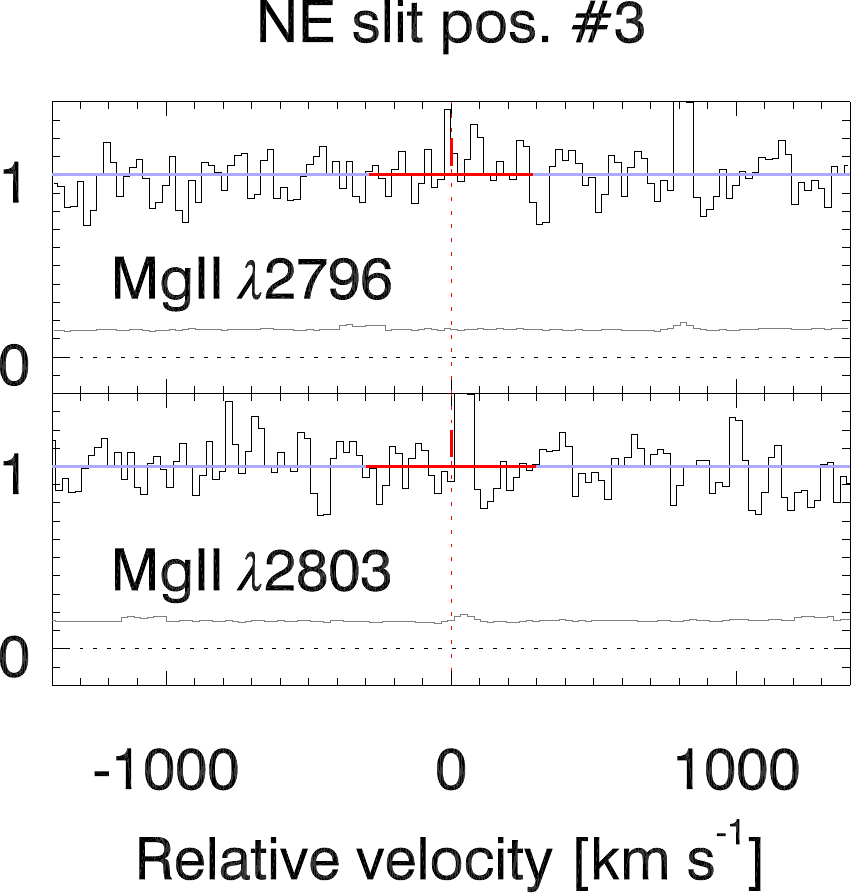}
\includegraphics[width=0.22\columnwidth,trim={0cm 0cm -0.3cm 0cm}]{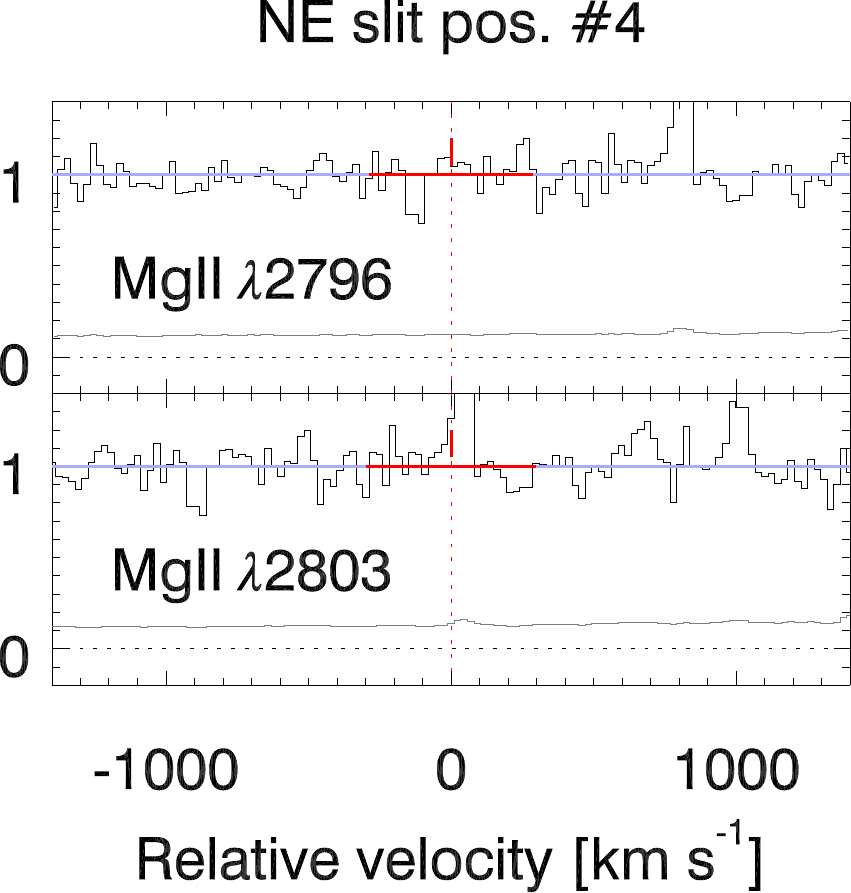}
\end{minipage}
\caption{Same as Fig.~\ref{voigt_profiles} for NE positions \#1 through \#4.}
\end{figure*}

\begin{figure*}
\begin{minipage}{2.00\columnwidth}
\includegraphics[width=0.22\columnwidth,trim={0cm 0cm -0.3cm 0cm}]{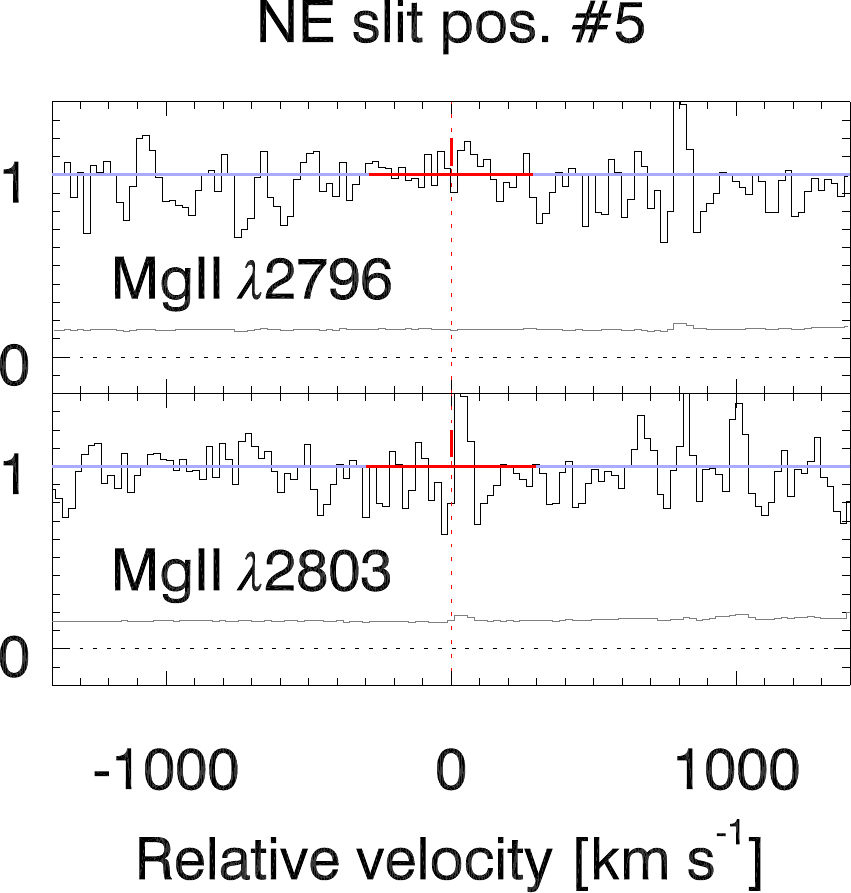}
\includegraphics[width=0.22\columnwidth,trim={0cm 0cm -0.3cm 0cm}]{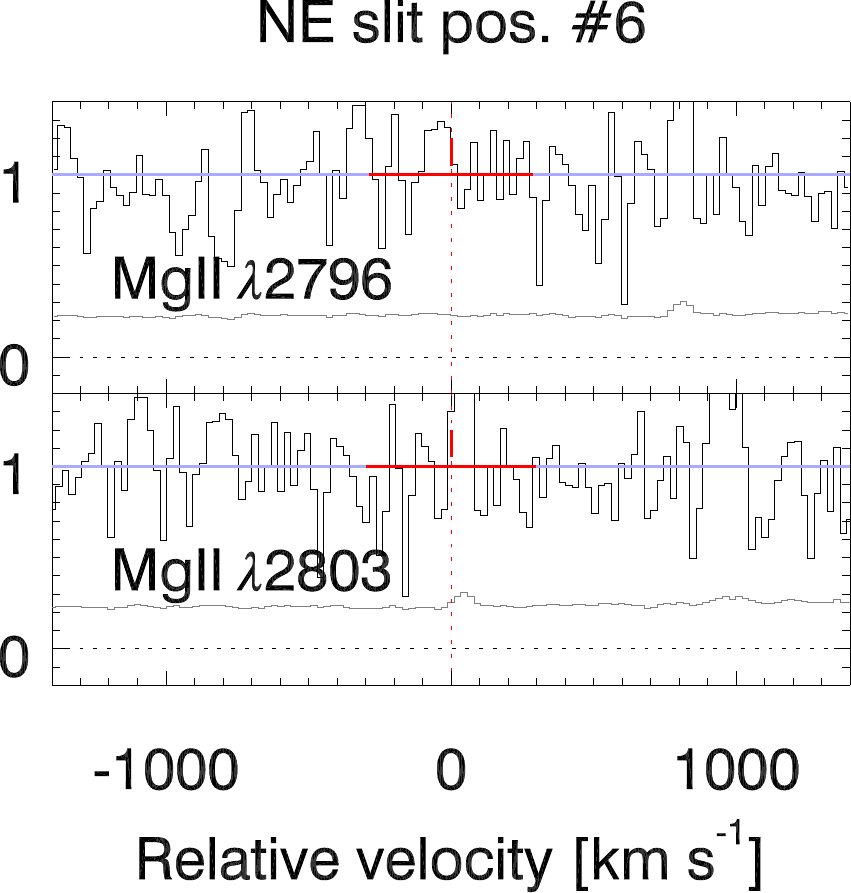}
\includegraphics[width=0.22\columnwidth,trim={0cm 0cm -0.3cm 0cm}]{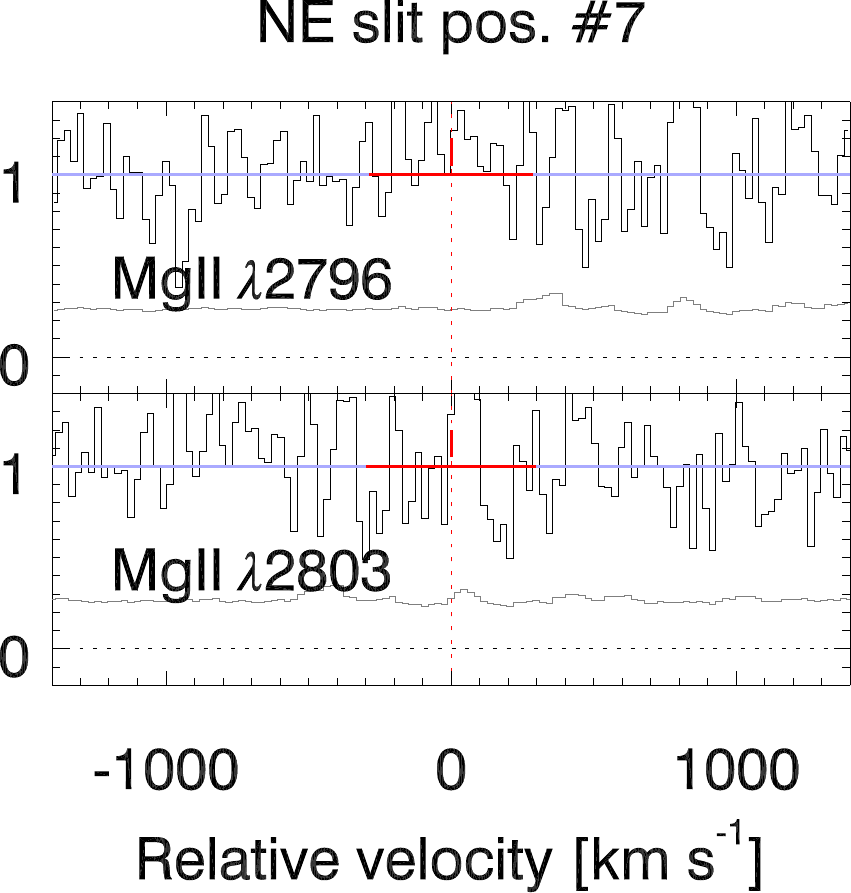}
\includegraphics[width=0.22\columnwidth,trim={0cm 0cm -0.3cm 0cm}]{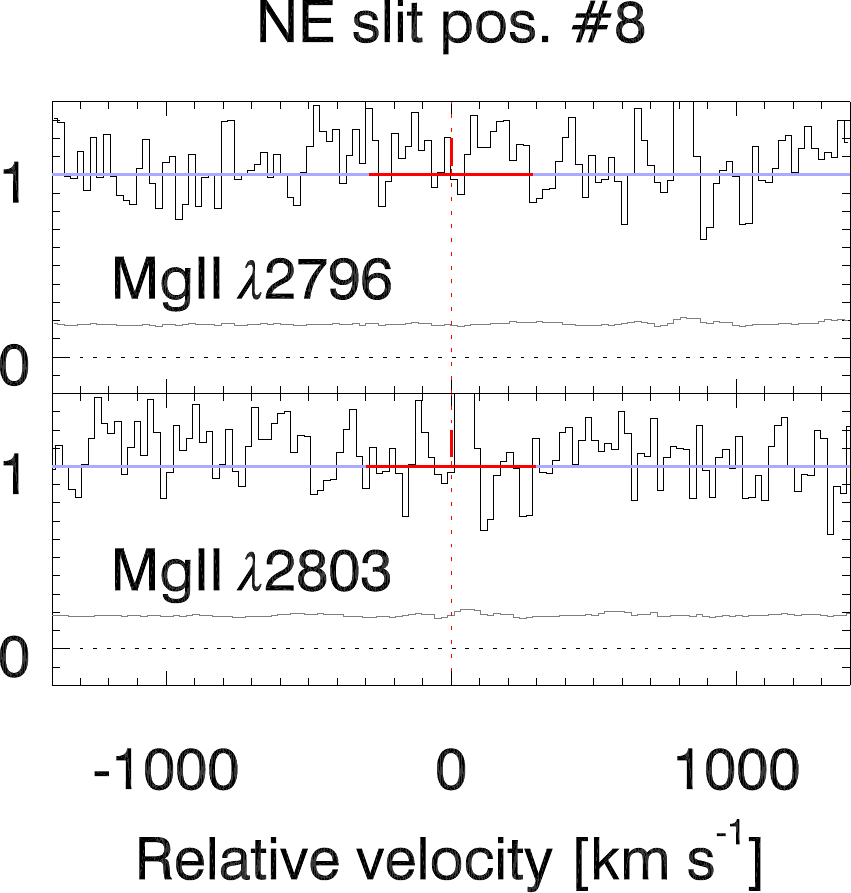}
\end{minipage}
\caption{Same as Fig.~\ref{voigt_profiles} for NE positions \#5 through \#8.}
\end{figure*}

\begin{figure*}
\begin{minipage}{2.00\columnwidth}
\includegraphics[width=0.22\columnwidth,trim={0cm 0cm -0.3cm 0cm}]{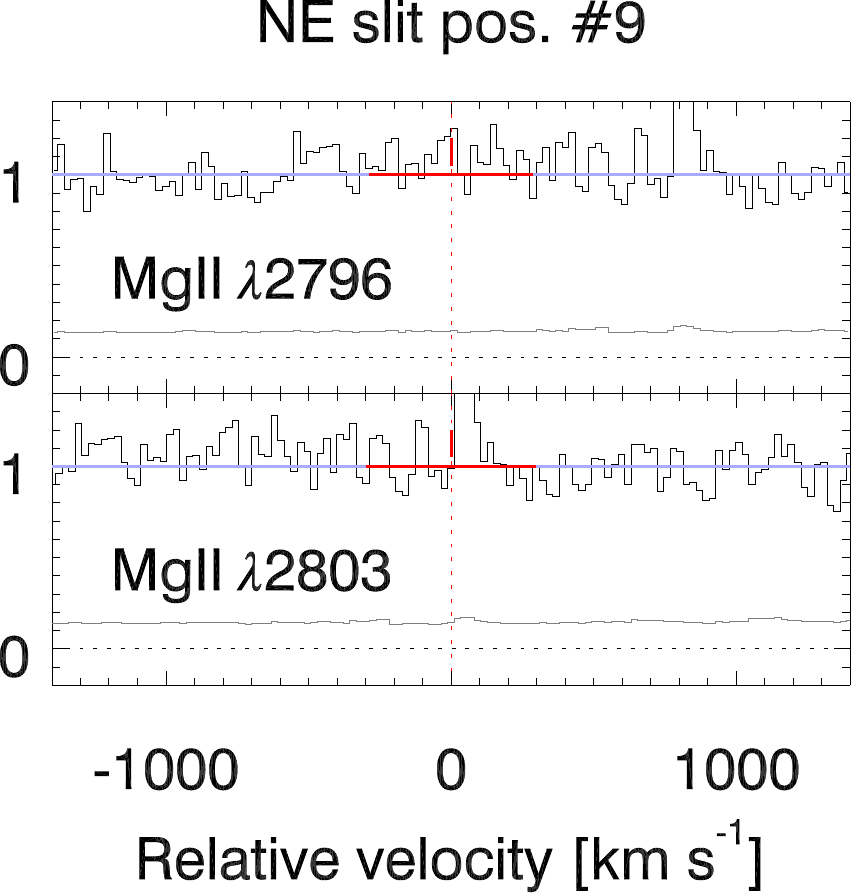}
\includegraphics[width=0.22\columnwidth,trim={0cm 0cm -0.3cm 0cm}]{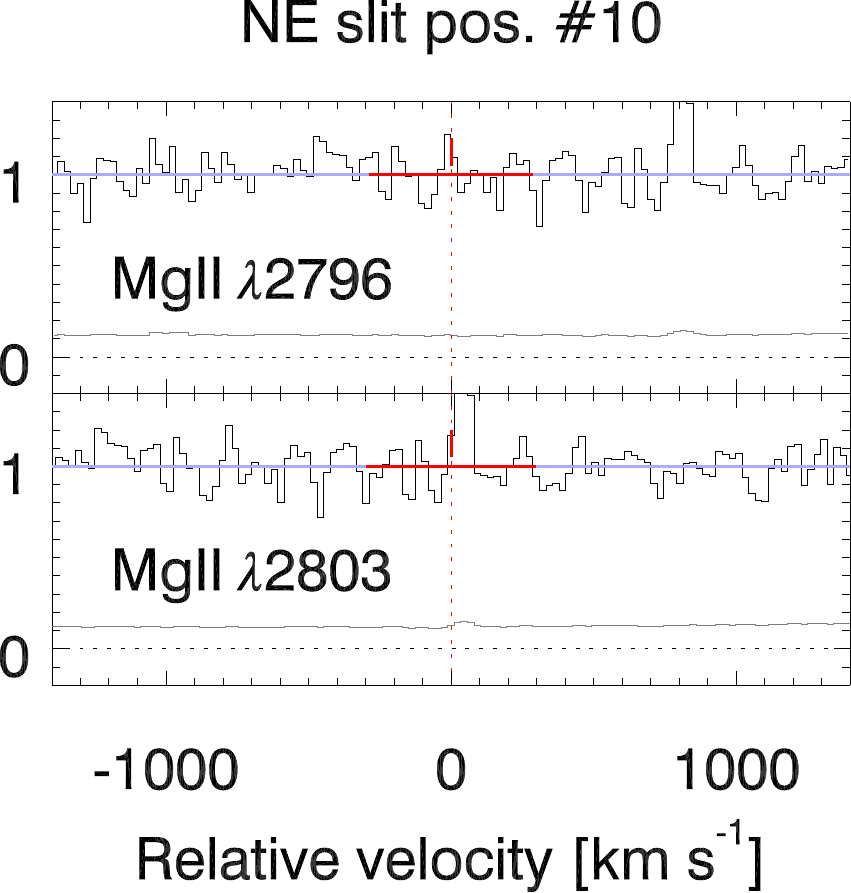}
\includegraphics[width=0.22\columnwidth,trim={0cm 0cm -0.3cm 0cm}]{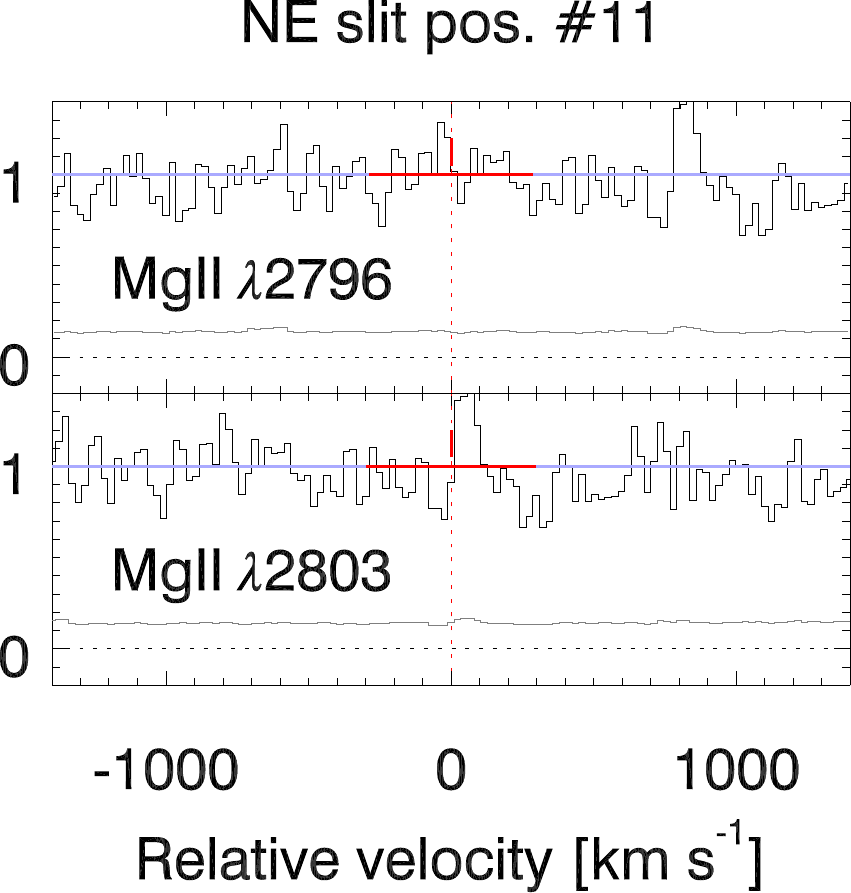}
\end{minipage}
\caption{Same as Fig.~\ref{voigt_profiles} for NE positions \#9, \#10 and  \#11.}
\label{voigt_profiles_last}
\end{figure*}

\end{document}